%% file: paper.tex
\documentclass[review]{acmsiggraph}

\TOGonlineid{0259}

\TOGvolume{0}
\TOGnumber{0}

\TOGarticleDOI{1111111.2222222}

\TOGprojectURL{}
\TOGvideoURL{}
\TOGdataURL{}
\TOGcodeURL{}

\graphicspath{{images/}}

\title{Efficient and Effective Volume Visualization with \\Enhanced Isosurface Rendering}

\author{}

\pdfauthor{}


\newcommand{\tabincell}[2]{\begin{tabular}{@{}#1@{}}#2\end{tabular}}

\begin{document}




\maketitle


\input{Abstract.tex}


\begin{CRcatlist}
  \CRcat{I.3.3}{Computer Graphics}{Picture/Image Generation}{Display algorithms};
  \CRcat{I.3.7}{Computer Graphics}{Three-Dimensional Graphics and Realism}{Color, shading, shadowing, and texture}
\end{CRcatlist}


\keywordlist


\TOGlinkslist


\copyrightspace


\input{Introduction.tex}

\input{RelatedWork.tex}

\input{OverallSystem.tex}

\input{ColorEnhancement.tex}

\input{Segmentation.tex}

\input{Experiments.tex}

\input{Discussion.tex}

\section*{Acknowledgements}

\bibliographystyle{acmsiggraph}
\bibliography{paper}
\end{document}

%% file: Abstract.tex
\begin{abstract}

Compared with full volume rendering, isosurface rendering has several well recognized advantages in efficiency and accuracy. However, standard isosurface rendering has some limitations in effectiveness. First, it uses a monotone colored approach and can only visualize the geometry features of an isosurface. The lack of the capability to illustrate the material property and the internal structures behind an isosurface has been a big limitation of this method in applications. Another limitation of isosurface rendering is the difficulty to reveal physically meaningful structures, which are hidden in one or multiple isosurfaces. As such, the application requirements of extract and recombine structures of interest can not be implemented effectively with isosurface rendering. In this work, we develop an enhanced isosurface rendering technique to improve the effectiveness while maintaining the performance efficiency of the standard isosurface rendering. First, an isosurface color enhancement method is proposed to illustrate the neighborhood density and to reveal some of the internal structures. Second, we extend the structure extraction capability of isosurface rendering by enabling explicit scene exploration within a 3D-view, using surface peeling, voxel-selecting, isosurface segmentation, and multi-surface-structure visualization. Our experiments show that the color enhancement not only improves the visual fidelity of the rendering, but also reveals the internal structures without significant increase of the computational cost.  Explicit scene exploration is also demonstrated as a powerful tool in some application scenarios, such as displaying multiple abdominal organs. 

\end{abstract}

%% file: Introduction.tex
\section{Introduction}

Isosurface rendering is one of the canonical techniques used for volume visualization. Compared with other visualization methods, ray-cast based isosurface rendering has several widely acknowledged advantages in efficiency and accuracy.  For example, the ray-isosurface intersection test can be very efficiently and accurately performed at the voxel level for each viewing ray. The empty space skipping techniques can also be intuitively implemented and yield very good performance. In comparison, some other volume rendering techniques such as semi-transparent rendering, also called full volume rendering, needs to take many samples in the volume, and involves intensive calculation. Although empty space skipping is also possible for the full volume rendering, the performance gain is relatively limited compared with isosurface rendering. 

\begin{figure}[!t]
\centering
    \begin{tabular}{cc}
    \includegraphics[height=1.8in]{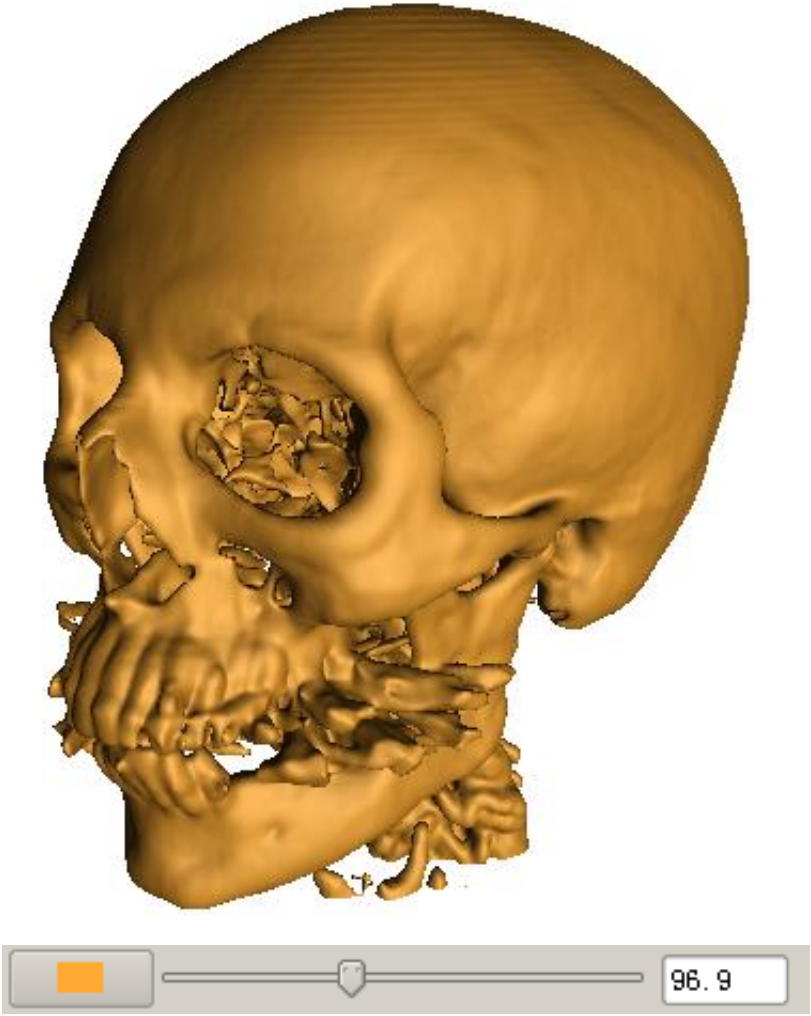}&
    \includegraphics[height=1.8in]{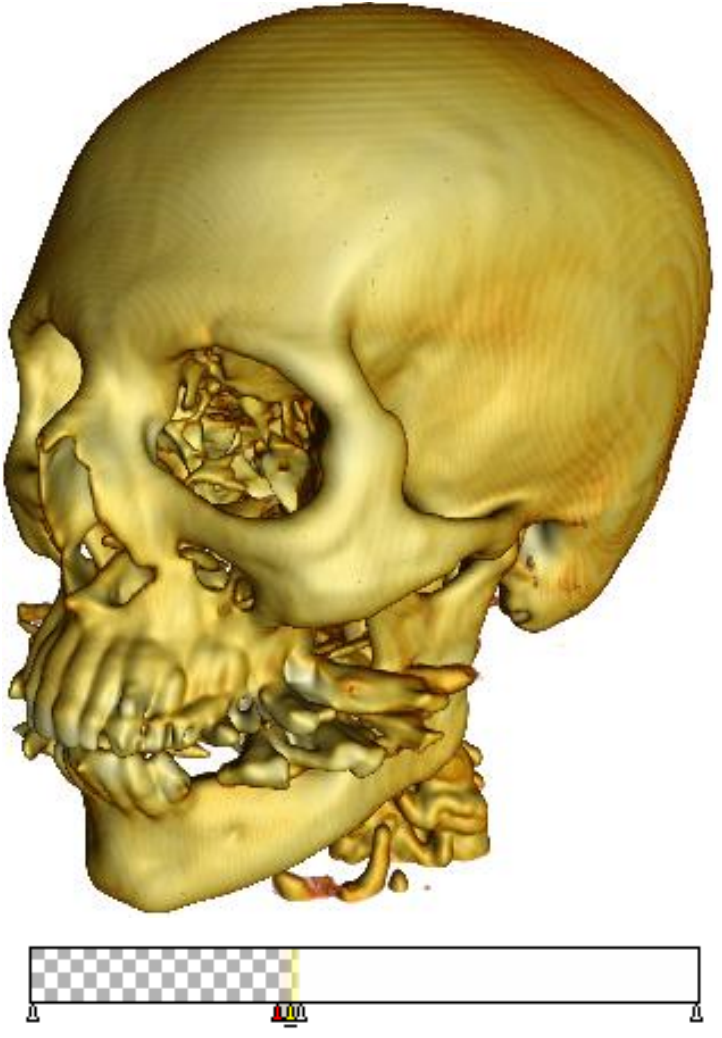}\\
    (a)&(b)\\
    \end{tabular}
\caption{Comparison of isosurface rendering and full volume rendering of the same isosurface. (a) Isosurface rendering. (b) full volume rendering. }
\label{Fig_Iso_vs_full}
\end{figure}

The standard isosurface rendering, however, also has some limitations in effectiveness. First, the isosurface rendering usually assigns a monotone color to an isosurface as the material color. Therefore, only the geometry properties of the isosurface can be visualized with shading. On the other hand, as we have noticed, full volume rendering is also capable of rendering isosurfaces, which can be done by using a transfer-function with a narrow transitional section. Visualizations achieved using such rendering setting can provide considerably more information about the neighborhood density, as shown in Figure \ref{Fig_Iso_vs_full}.  In many applications, it can be significantly beneficial to visualize the internal structures behind an isosurface. For example, in CT colonography, clinicians want to identify the nature of polyps. But from a monotone colored isosurface rendering, there's no information of the internal structure for decision making. To address this requirement, \cite{Pickhardt:2004:translucency} conducted a full volume rendering with a selected window so that the internal structure is revealed. However, additional computation is needed and the overall performance will be hurt. 

Another limitation of the isosurface rendering is the difficulty to identify structures of interest and to efficiently visualize multiple structures of interests in a single view. Usually, an isosurface contains many different physically meaningful structures, and the users are interested with a few of them. The rest of the structures, as they are connected as a whole, are not only unnecessary, but also visually confusing. To tackle this problem, segmentation can be considered. However, direct volume segmentation methods usually do not integrate very well with isosurface rendering, and artifacts are likely to be introduced. 

In this work, we develop an enhanced isosurface rendering technique to improve the effectiveness while maintaining the performance efficiency of the standard isosurface rendering. First,  we propose an {\em  isosurface color enhancement method}  to illustrate the material property and the internal structures behind the isosurface. Our approach considers the neighborhood volume directly behind the isosurface, and applies this information to derive the property of local material varying along the isosurface. Since it is a coloring/shading method, the voxel-traversal and ray-voxel intersection scheme is not changed from the original isosurface rendering. Therefore the simplicity of the standard isosurface rendering is kept the same, and computational complexity of the enhanced isosurface rendering is not increased.  Second, we enhance the structure extraction capability of isosurface rendering by enabling {\em explicit scene exploration} within a 3D-view, using surface peeling, voxel-selecting,  isosurface segmentation, and multi-structure visualization. During this process, different structures can be extracted from different isosurfaces intuitively. Then, the surface structures are recombined into a single scene for display. The segmentation and recombination is done within the subset of the voxels that contain the required isosurface, so the segmentation result is very suitable to be integrated with the isosurface rendering scheme.

The remainder of this paper is organized as follow. In Section 2, we introduce the related works in volume rendering. In Section 3, we provide an overview of our proposed approach. In Section 4, we describe the model and implementation of the isosurface color enhancement. In Section 5, we illustrate the techniques used in the explicit scene exploration framework. The experiment results are discussed In Section 6. In Section 7, we discuss some related issues and the future development of this work.

%% file: RelatedWork.tex
\section{Related Work}

\subsection{Isosurface Rendering}

Isosurface rendering has been intensively studied over the years, and many techniques has been developed which makes isosurface rendering a highly efficient and accurate approach for volume visualization. To identify the intersection of a viewing ray and an implicitly defined isosurface in a volume, two techniques are involved. One is ray traversal, the other is intersection test. 

For ray traversal, two typical schemes are uniform sampling and voxel oriented traversal. Uniform sampling is more widely used in GPU based volume ray-casting \cite{Stemaier:2005:SFVR}, and it is preferred when used along with some algorithms like pre-integration \cite{engel:2001:HPV}\cite{lum:2004:HLA}. On the other hand, voxel oriented traversal provides an opportunity to do a precise intersection test. In earlier volume ray-casting techniques designed for CPU, the 3DDDA algorithm \cite{Amanatides:1987:AFV} is often used, such as in \cite{Parker:1998:IRT}. 

There are also a couple of ways to perform the intersection test. The simplest way is to base on a linear approximation between samples. An equivalent method is to use a pre-integration. In voxel oriented ray traversal schemes, if tri-linear interpolation is used, the value along the ray segment within a voxel can be considered as a cubic function, the intersection test is converted to a cubic equation problem, which can be solved in a closed form \cite{Parker:1998:IRT} or using piecewise recursive root finding \cite{Marmitt:2004:FAA}. Approximate methods can provide a faster but inaccurate result as is given in \cite{Neubauer:2002:CFR} and Scharsach2005\cite{Scharsach:2005:AGR}. 

When GPU is used, tri-linear interpolation is supported by the hardware. In that case, performance can be further improved by fetching fewer samples. In \cite{Ament:2010:DIVV}, it is pointed out that as few as 4 data fetches per voxel is enough for the coefficient extraction. In this work, we applied a similar strategy for the ray-isosurface intersection test. 

\subsection{Full Volume Rendering}

The core concept of full volume rendering is volume rendering integral, which is based on an optical model described in Max~\cite{Max:1995:OMF} The integration is performed for a viewing ray cast from the viewpoint, where $l = 0$, to the far end, where $l = Far$.  $I_0$ is the light coming from the background, $\mu$ is the per-unit-length extinction coefficient, and $c$ is a intensity or color. In this paper, the neighborhood of the isosurface to be rendered is considered in this model. 

\[
\begin{array}{l}
I = {I_0}\exp \left( { - \int_0^{Far} {\mu \left( l \right)dl} }
\right) \\
+ \int_0^{Far} {c\left( l \right)\mu \left( l \right)\exp \left( { -
\int_0^l {\mu \left( t \right)dt} } \right)dl}
 \end{array}
\]

To distinguish different contents in a volume dataset during rendering, classification and segmentation are two commonly used technologies. Classification can be done by a variety of different transfer-functions, which are different from each other mainly in the feature domains where they are defined. While the most commonly used transfer-function is the intensity base 1D-transfer-function, there are loads of other transfer-functions which are based on other features such as gradient magnitude ~\cite{kniss:2002:MTF},  size feature ~\cite{correa:2008:size}, and texture feature ~\cite{caban:2008:texture}. Although efforts have been made, classification based methods are just partially capable of the content distinguishing task. Sometimes, segmentation is necessary. As another intensively studied area, most segmentation methods are not specifically designed for visualization, and the integration of segmentation results and volume ray-casting has been problematic.

To visualize a segmented volume, different transfer functions can be assigned to different segments of the volume. An issue has to be dealt with is the boundary filtering. On vector GPUs, a solution is given in ~\cite{hadwiger:2003:HTV}. However, most recent GPUs use scalar architecture for better programmability, in which case, another solution given in ~\cite{xiang:2010:SCE} is more suitable. In this work, instead of volume segmentation, we use isosurface segmentation, which works on the subset of voxels containing the isosurface. 

\subsection{Min-Cut Algorithm}

Many segmentation issues can be converted into a min-cut question in graph theory. The optimization method described by ~\cite{boykov:2004:exp} provides a very efficient tool to deal with this kind of questions. In our work, we define the isosurface segmentation as a min-cut problem and use this algorithm for the optimization.

%% file: OverallSystem.tex
\section{Overview}

\begin{figure*}[!t]
\centering
\includegraphics[width=6.0in] {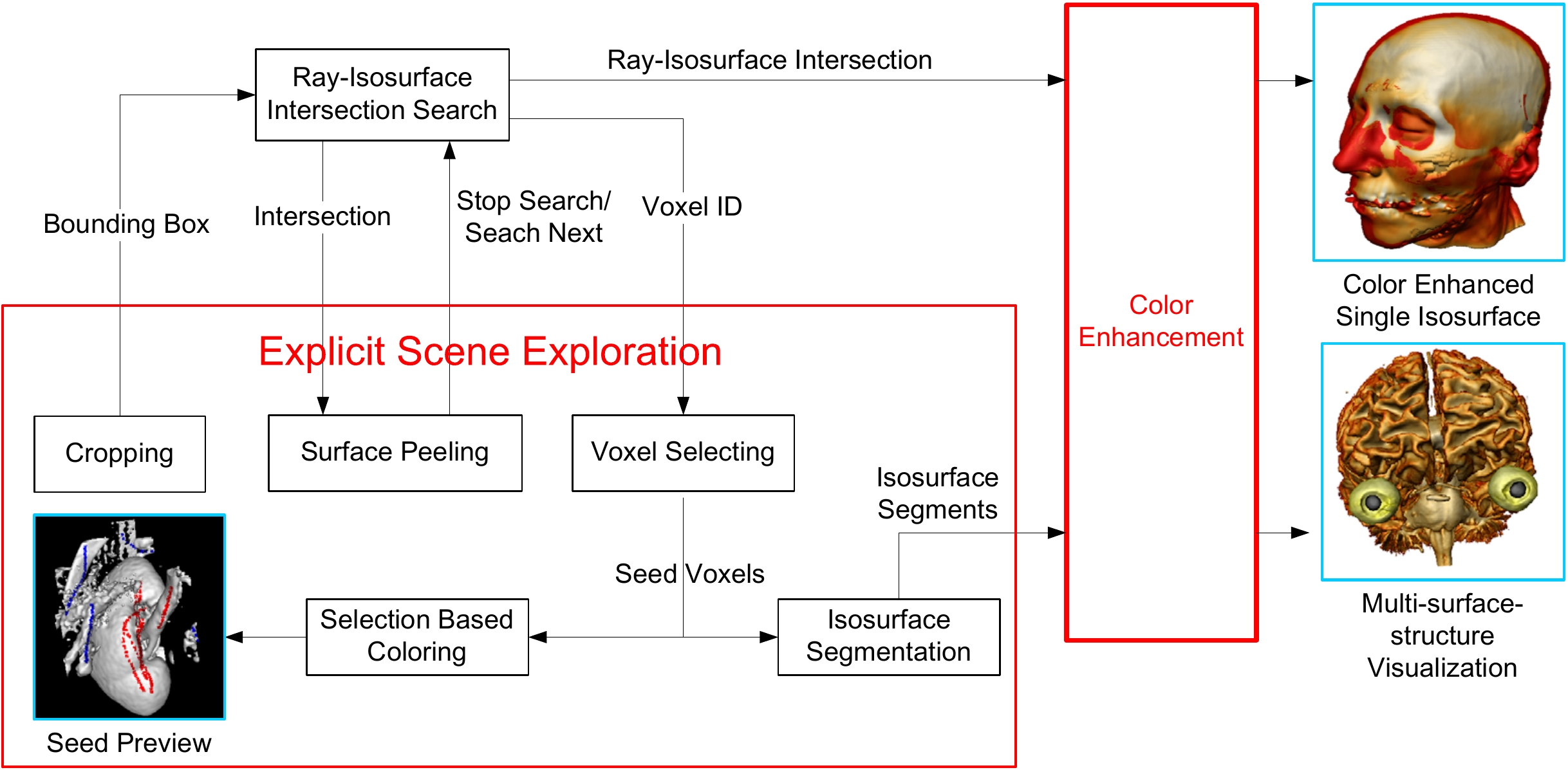}
\caption{The overview of the contributions of our approach. The proposed enhanced isosurface rendering technique is accomplished by the color enhancement method and the explicit scene exploration scheme. Explicit scene exploration consists of surface peeling,  voxel-selecting, isosurface segmentation, and multi-surface-structure visualization.  } \label{Fig_Overall}
\end{figure*}

In this section, we provide an overview of our approach, as is shown in
Figure~\ref{Fig_Overall}, mainly focusing on the contributions of this work. 

The first contribution, the {\bf color enhancement technique}, aims to visualize the neighborhood density and reveal the internal structures behind the isosurface. The technique is applied to each of the intersection points on the isosurface to provide a material color. Our second contribution, the {\bf explicit scene exploration scheme} is a composition of several novel techniques
aiming to enhance the structure extraction capability of isosurface rendering. When provided with a volume dataset which we have little knowledge, cropping and surface peeling can be used for a brief exploration so that the structures of interest can be quickly identified. Then, using the isosurface segmentation, the structures of interest can be extracted and represented as isosurface segments. With a slightly modified isosurface intersection search procedure along with the color enhancement technique, the isosurface segments can be rendered within a single scene, which we call the {\bf multi-surface-structure visualization}. During the isosurface segmentation, a few seed voxels are specified by the user. An intuitive seed selecting interface is designed to handle this task, which involves two techniques: voxel selecting and selection based coloring. While in voxel selecting, seed points are collected by user's mouse clicks and drags on the image plane. The selection based coloring provides a preview of these seeds. The interaction process works directly in a 3D view, that's why we call it ``explicit scene exploration''. During this process, the voxel based ray-traversal and ray-isosurface intersection search is an essential part of the system, which yields an accurate ray-isosurface intersection for each ray. The IDs of the voxels containing the intersection points are also collected in this process.

%% file: ColorEnhancement.tex
\section{Color Enhancement}

Traditional isosurface rendering usually assigns a monotone color to an isosurface as the material color. The reason is that every point belonging to an isosurface has the same scalar value. Since the only scalar value is used for defining the geometry of the structure, there's no information directly available to define the material of that surface. However, if we consider a small neighborhood directly behind the isosurface, it is possible to get some information of the of local material varying along the isosurface. 

\subsection{Visualization Model}

\begin{figure}[!t]
\centering
\includegraphics[width=2.5in] {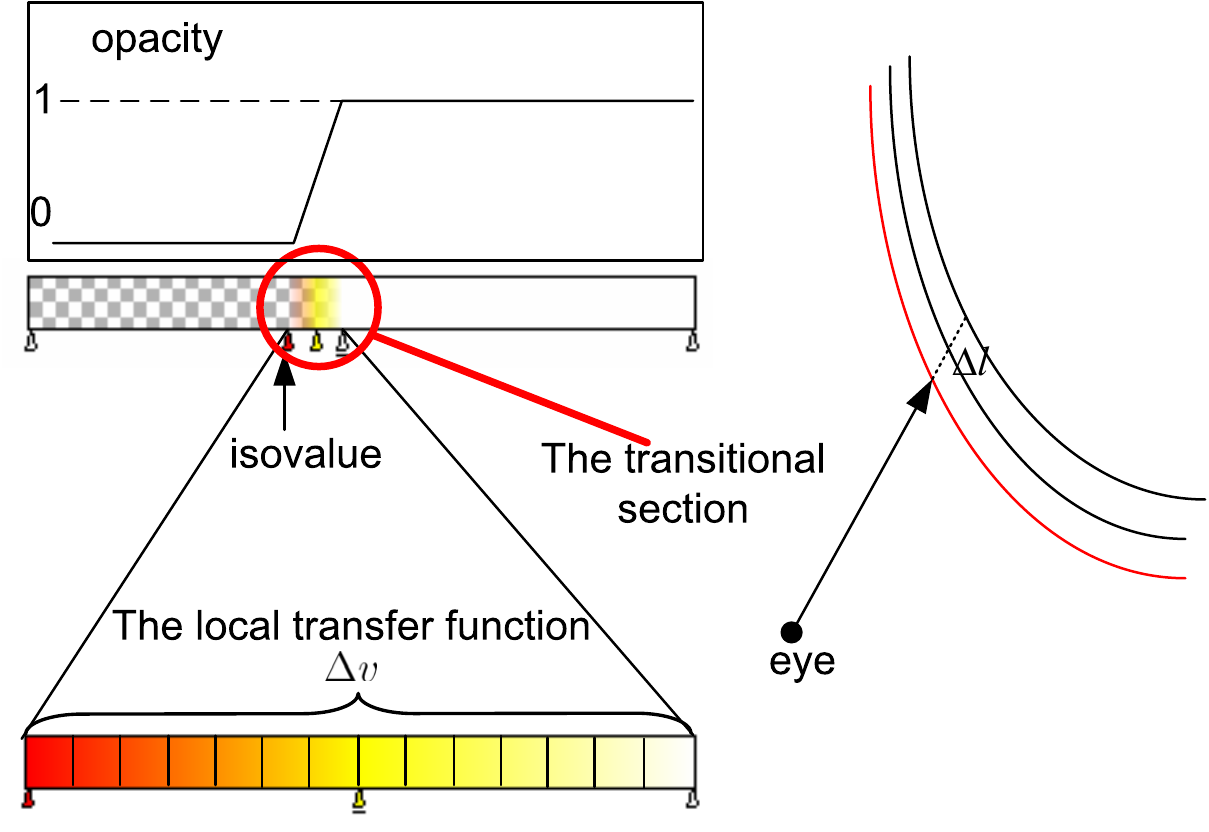}
\caption{Transfer-function defined by a single transitional section} 
\label{Fig_TF}
\end{figure}

Inspired by full volume rendering, we consider the isosurface neighborhood to be semi-transparent. In full volume rendering, such a neighborhood can be defined by a transfer-function in which the opacity transits from 0 to 1 within a narrow range near the isovalue. As shown in Figure \ref{Fig_TF}, such a transitional section can be specified by an isovalue and a local transfer function. The transitional section ranges from $isovalue$ to $isovalue+\Delta v$. Scalar values less than $isovalue$ are mapped to fully transparent, and scalar values greater than $isovalue+\Delta v$ are mapped to fully opaque. 

If the neighborhood of the isosurface corresponding to $\Delta v$ is sufficiently thin, it can be assumed that most of the viewing rays hitting the isosurface are going through the whole transitional range and the alpha value finally accumulates to 1. In this case, the accumulated color is mainly decided by the thickness of the neighborhood $\Delta l$ where the accumulation happens. If $\Delta l$ is relatively big, the color of the nearer neighborhood plays an important role, and the contribution of the farther neighborhood is relatively small since the opacity accumulates. Reversely, if $\Delta l$ is small, the color of the farther neighborhood will count more. 

\begin{figure}[!t]
\centering
\includegraphics[width=2.5in] {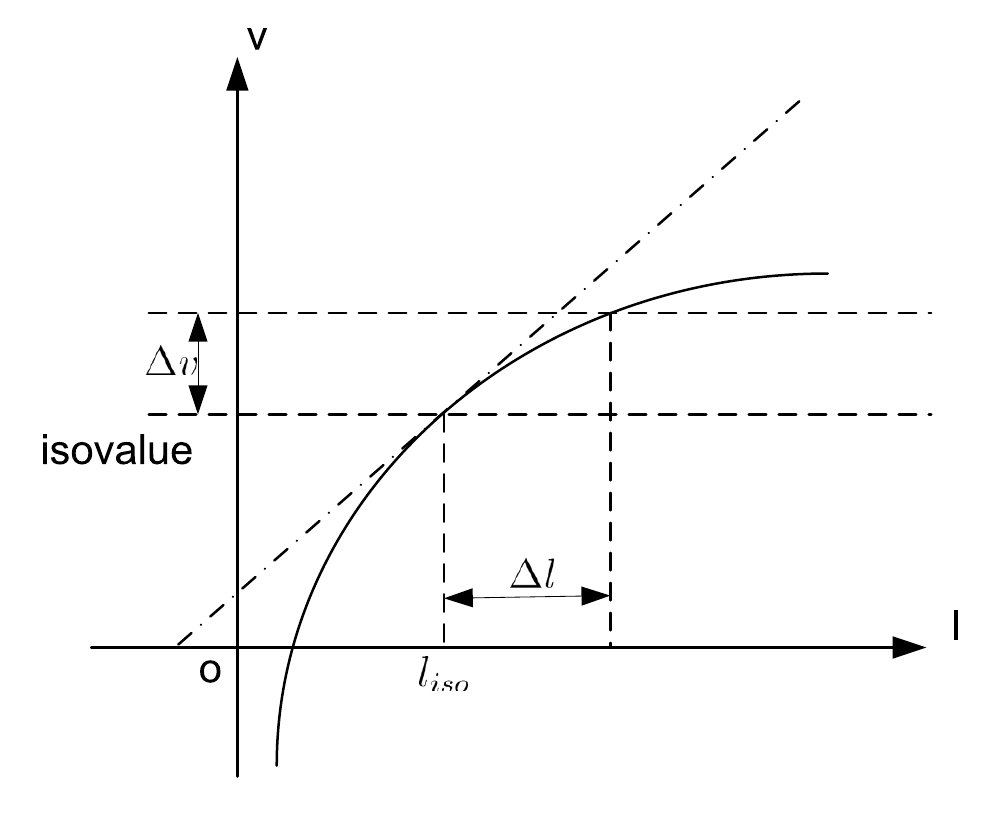}
\caption{Linear Approximation of the Isovalue Neighborhood. $l$ denotes the length of ray. $v$ denotes the scalar value.} 
\label{Fig_LinearApprox}
\end{figure}

From the intuitive observation, we consider to use the scalar changing rate (or directional derivative) as a local material hint, since $\Delta l$ can be linearly approximated by dividing the value range $\Delta v$ by the directional derivative. 

However, let's still begin with $\Delta l$, because it is directly related to the alpha values used for accumulation in full volume rendering. In full volume rendering, if $n$ scalar values are evenly sampled within the isosurface neighborhood of thickness $\Delta l$,  the alpha value of a sample used for accumulation can be calculated as:

\[\alpha  = {\alpha _T}^{\frac{{{{\Delta l} \mathord{\left/
 {\vphantom {{\Delta l} n}} \right.
 \kern-\nulldelimiterspace} n}}}{{stdSampleDistance}}}\]
 
, where $\alpha _T$ denotes the alpha value defined in the transfer-function assuming a sample distance $stdSampleDistance$. With linear approximation, there is:

\[\begin{array}{l}
 \Delta l \approx \frac{{\Delta v}}{{{{v'}_l}\left( {{l_{iso}}} \right)}} \\ 
 \end{array}\]

, where ${{v'}_l}\left( {{l_{iso}}} \right)$ denotes the scalar changing rate at the intersection point. So:

\begin{equation}
\alpha  \approx {\alpha _T}^{\frac{{{{\Delta v} \mathord{\left/
 {\vphantom {{\Delta v} n}} \right.
 \kern-\nulldelimiterspace} n}}}{{stdSampleDistance{{v'}_l}\left( {{l_{iso}}} \right)}}}
 \label{ApproxAlpha}
\end{equation}

Now, we define:

\[\begin{array}{l}
densityFactor = \frac{{\Delta v}}{{stdSampleDistance}}\\
speed=\frac{{{{v'}_l}\left( {{l_{iso}}} \right)}}{{densityFactor}}, speed \in \left( {0, + \infty } \right)
\end{array}\]

In this way, Equation (\ref{ApproxAlpha}) can be writen as: $\alpha  \approx {\alpha _T}^{\frac{1}{{n \cdot {\rm{speed}}}}}$

From the above formulations, we know that:

\begin{itemize}
\item A value $speed$ can be easily calculated from the scalar changing rate at the intersection point, ${{v'}_l}\left( {{l_{iso}}} \right)$, with $densityFactor$ as a global tunable parameter.
\item Given the value $speed$, the alpha values used for accumulation can be approximated.
\end{itemize}

\subsection{Speed-color Map}

During rendering, we use  ${{v'}_l}\left( {{l_{iso}}} \right)$ to characterize the local material. To efficiently calculate the material color from the ${{v'}_l}\left( {{l_{iso}}} \right)$, a map from $speed$ to color $\vec C$ is built as follow:

Given a local transfer-function 
\[\left\{ {{\alpha_{Ti}}, {{\vec c}_i} = \left( {{r_i},{g_i},{b_i}} \right),i = 1,2,3,...,n} \right\}\]

, for each $speed$, ${\alpha _{speed,i}} = {\alpha _{Ti}}^{\frac{1}{{n \cdot {\rm{speed}}}}}$. The accumulated color 
$\vec C\left( {{\rm{speed}}} \right)$ can be pre-calculated by an alpha blending:

\[\vec C\left( {{\rm{speed}}} \right) = \sum\limits_{i = 1}^n {{\alpha _{speed,i}}{{\vec c}_i}} \prod\limits_{j = 1}^{i - 1} {\left( {1 - {\alpha _{speed,j}}} \right)} \]

However, since $speed$ has an infinite value range, we cannot simply build a speed-color map linearly. In our implementation, we use a logarithm sampled speed-color map of size $m$,  $\left\{ {{{\vec C}_j} = \vec C\left( { - \ln \left( {1 - {\raise0.7ex\hbox{$j$} \!\mathord{\left/ {\vphantom {j m}}\right.\kern-\nulldelimiterspace} \!\lower0.7ex\hbox{$m$}}} \right)} \right),j = 1,2,3, \cdots ,m} \right\}$ to store the pre-calculated material colors. 

\subsection{Estimation of Scalar Changing Rate}

\begin{figure}[!t]
\centering
    \begin{tabular}{ccc}
    \includegraphics[width=1.0in]{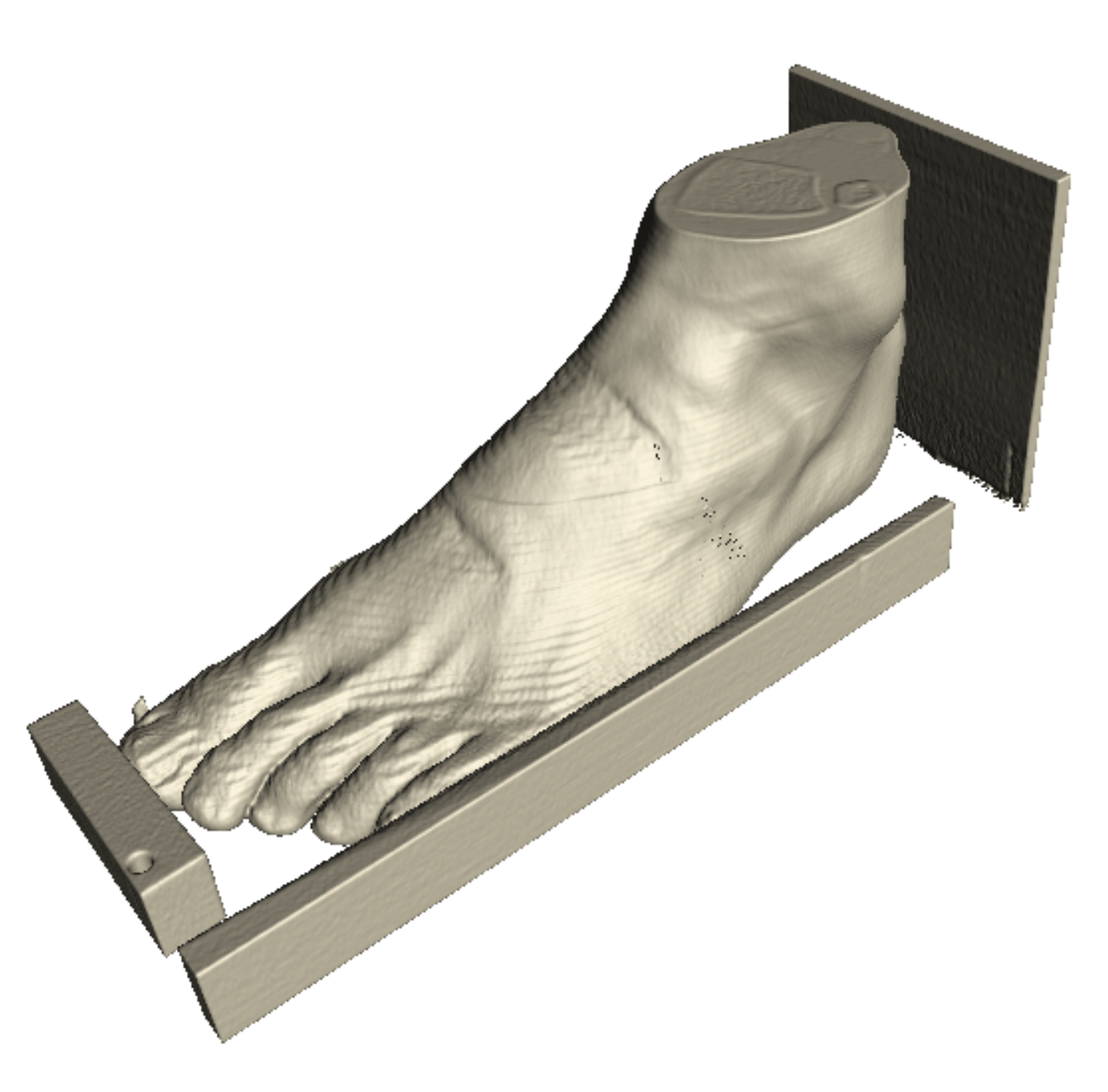}&
    \includegraphics[width=1.0in]{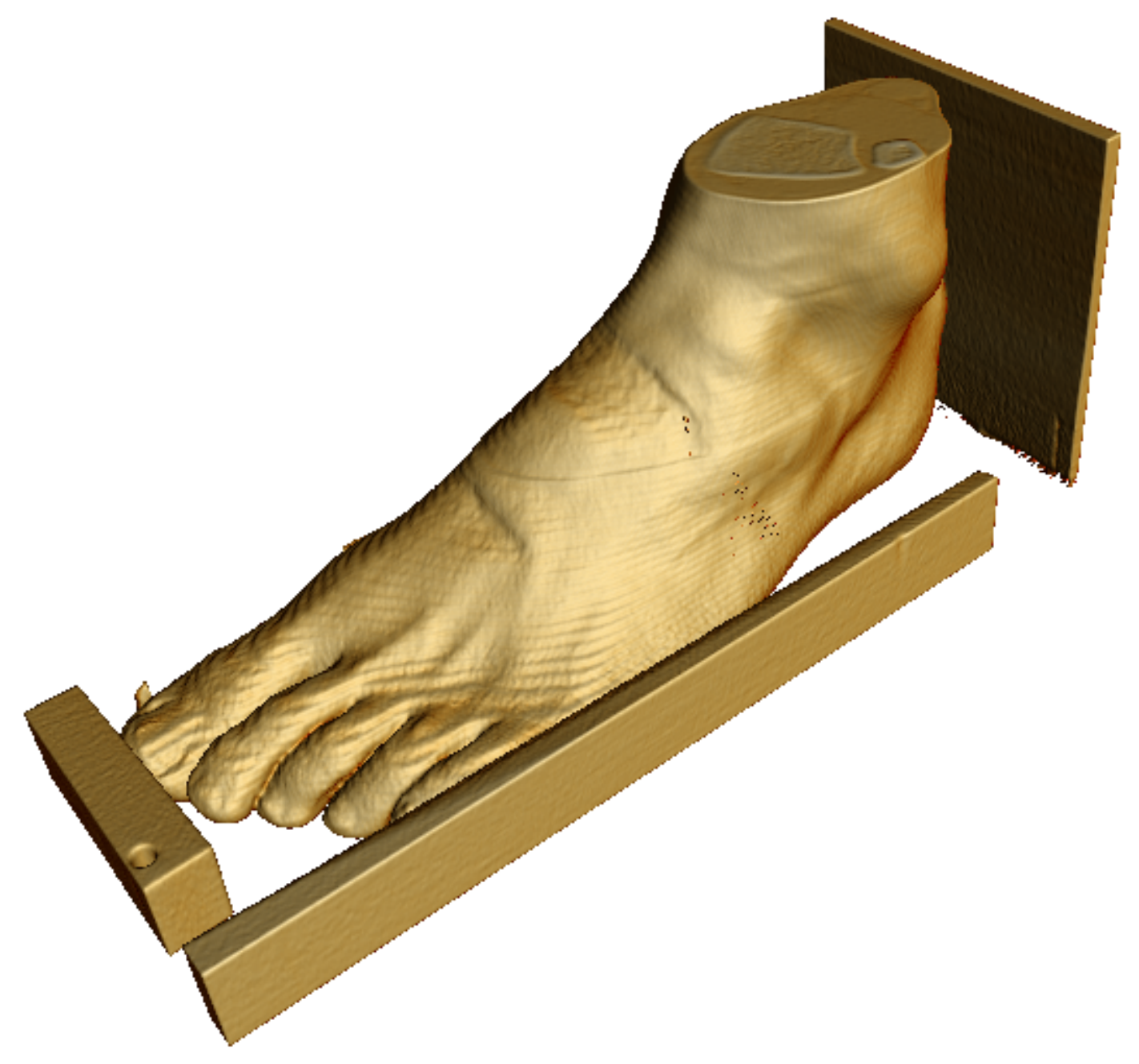}&
    \includegraphics[width=1.0in]{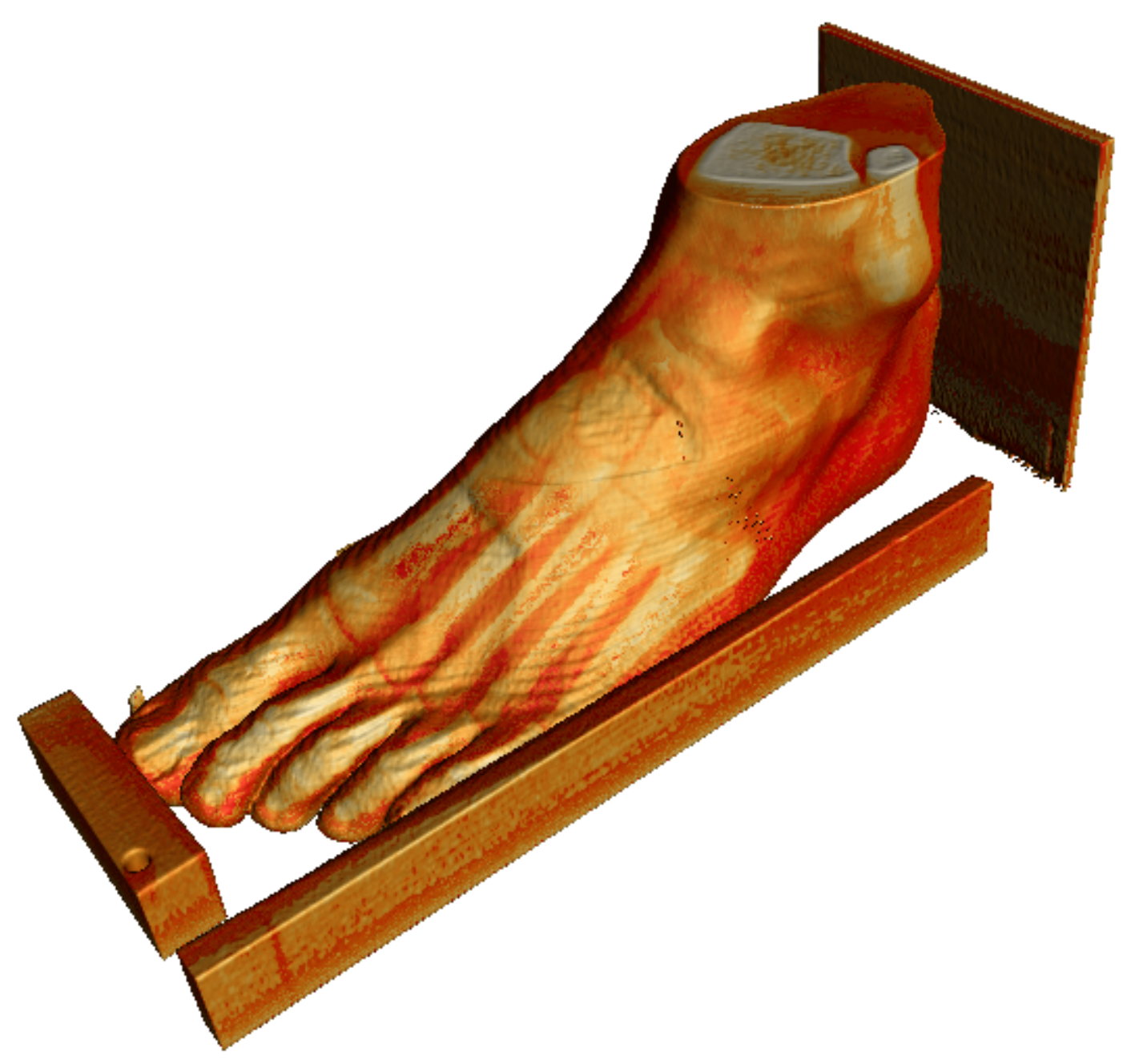}\\
    (a)&(b)&(c)\\
    \end{tabular}
\caption{ Displaying shallow and deep neighborhood with different estimations of the scalar changing rate. (a) Mono colored. (b) Displaying shallow neighborhood, estimated with the gradient vector. (c) Displaying internal structures. }
\label{Fig_Shallow_And_Deep}
\end{figure}

The scalar changing rate ${{v'}_l}\left( {{l_{iso}}} \right)$ can be estimated in different ways during rendering, which has a significant effect on the rendering result.  For basic coloring need, we can simply use $\nabla v \cdot \vec r$, where $\vec {\nabla} v$ denotes the gradient vector, which will be calculated for shading anyway, and $\vec r$ is the direction vector of the viewing ray. In our implementation, $\vec {\nabla} v$  is calculated on the fly with 6 neighboring scalar samples. However, using an estimation like this, only a shallow neighborhood behind the isosurface can be reflected by the resulted color, as Figure \ref{Fig_Shallow_And_Deep} (b) shows. 

To reveal the internal structures, scalar changing rate should be estimated within a larger scope. To this end, we add in $\Delta v$ as another global tunable parameter. In scalar changing rate estimation, we first perform an iterative search along the viewing ray to find a $\Delta l$ where the scalar value reaches $isovalue + \Delta v$. Then the scalar changing rate is estimated by $\Delta l / \Delta v$. In this way, we find that many of the internal structures can be revealed without noticeable performance impact. Figure \ref{Fig_Shallow_And_Deep} (c) is an example, and more will be presented in the result part.

%% file: Segmentation.tex
\section{Explicit Scene Exploration}

An innate advantage of isosurface rendering is that, for each ray, a definite intersection point can be generated. When voxel based ray-traversal is used, the exact voxel can also be identified. This feature can be very useful in interaction design. For example, it enables the direct picking of voxels from the volume by mouse clicking and dragging within the 2D image plane. By exploiting this advantage, we develop an explicit scene exploration framework to quickly identify the structures of interest, pick them out, and recombine them into a single scene. 

\subsection{Accuracy Assurance}

To guarantee the quality of the interaction, such as the voxel selecting operation, it is crucial to make sure that the ray-isosurface intersection test is accurate for each ray. First, voxel based ray traversal such as the 3DDDA algorithm should be used. Second, the intersection finding within the voxel should be accurately performed. 

\begin{figure}[!t]
\centering
\includegraphics[width=1.2in] {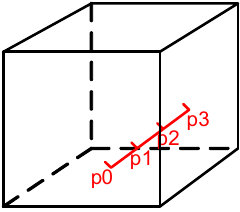}
\caption{Use four values along the ray in each voxel to estimate the coefficients of the cubic polynomial.} 
\label{Fig_samples}
\end{figure}

When trilinear interpolation is assumed, the value along the ray segment within each voxel can be expressed as a cubic function to the ray parameter. Once the coefficients of the cubic function are decided, the intersection test can be transformed into solving a cubic equation, which can be efficiently solved using the method proposed by Marmitt et al. ~\cite{Marmitt:2004:FAA}. There are different ways to decide the coefficients of the cubic function. We choose to use four values along the ray in each voxel to estimate the coefficients of the cubic polynomial, as is shown in Figure \ref{Fig_samples}. The viewing ray enters the voxel from $p0$ and exits from $p3$. A value is fetched at $p0$, $p1$, $p2$, and $p3$ each respectively, where $p1$ and $p2$ are the trisection points of the ray-voxel intersection. 

\subsection{Surface Peeling}

When provided with a volume data which we have little knowledge, the first issue of visualization should be to explore the dataset and identify the structures of interest. This has not been a simple issue for both full volume rendering and monotone isosurface rendering, due to the fact that many of the structures overlap each other. 

\begin{figure}[!t]
\centering
    \begin{tabular}{cc}
    \includegraphics[width=1.5in]{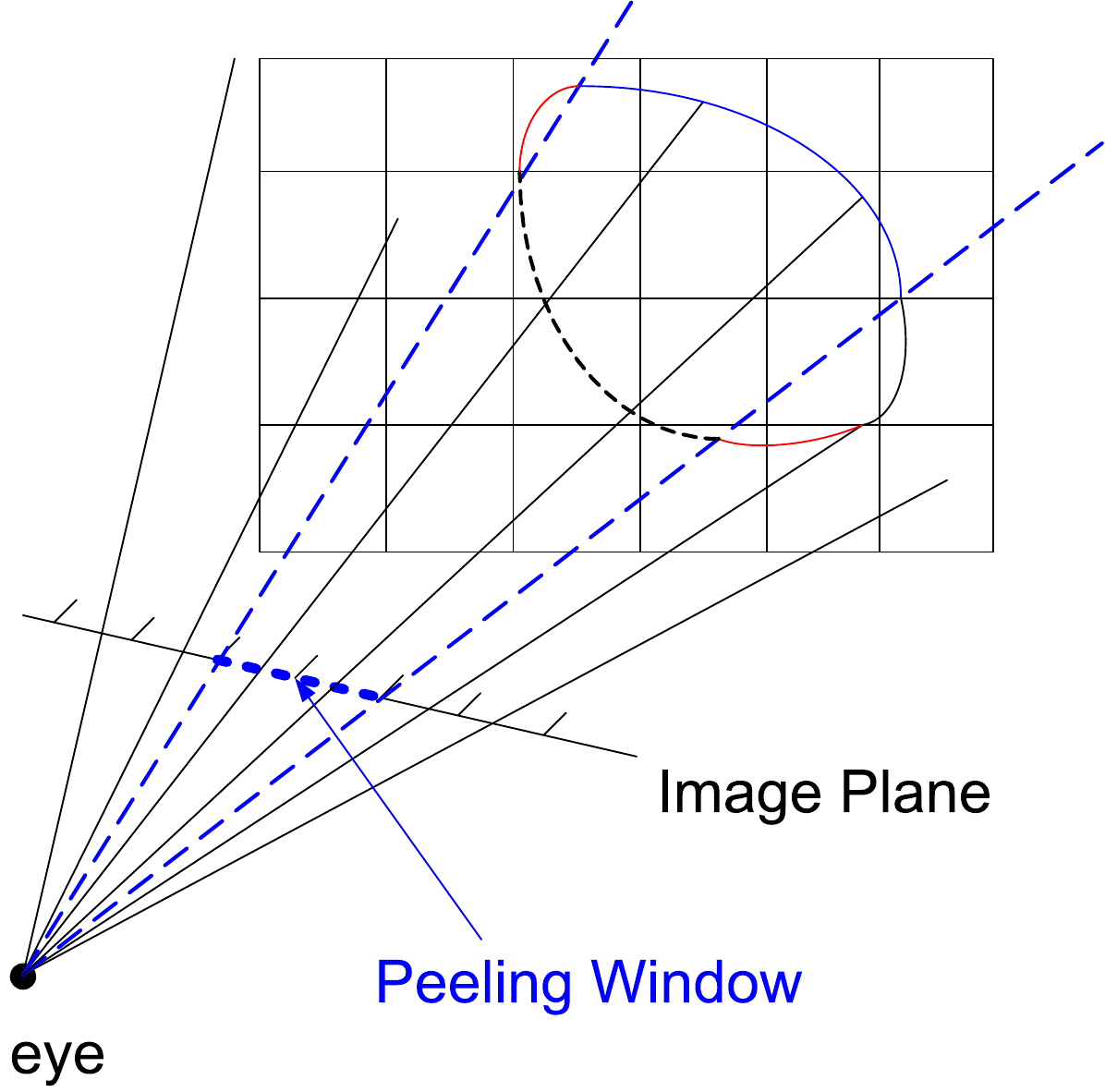}&
    \includegraphics[width=1.5in]{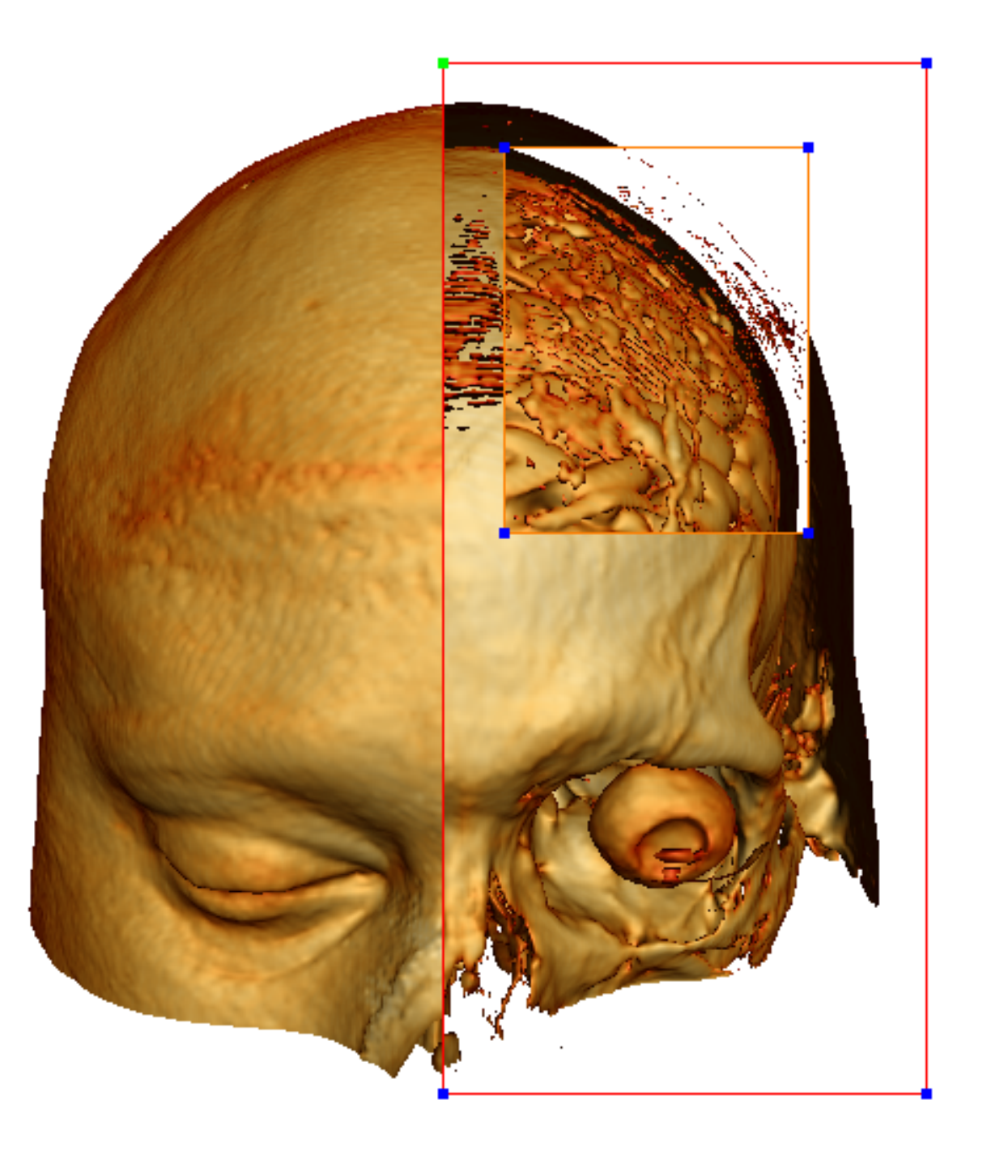}\\
    (a)&(b)\\
    \end{tabular}
\caption{Surface Peeling. (a) Illustration of surface peeling. (b) An example of surface peeling. } 
\label{Fig_peeling}
\end{figure}

For isosurface rendering, we can use cropping and surface peeling for the brief exploration. Cropping is a simple technique which just trunk the volume with a bounding box during rendering to localize the rendering region of the volume. Surface peeling is performed by jumping over the first few intersection points within some selected image plane areas, so the the structures behind can be exposed. As is shown in Figure \ref{Fig_peeling} (a), a peeling window is defined on the image plane. Within that window, the first ray-isosurface intersection is skipped over, so the second layer of the isosurface is rendered. In our implementation, we use an integer array as the "peeling buffer". The values are initialized as 0. Within each peeling window, the peeling value increases by 1. During rendering, the number of intersections is counted, if the count is no-bigger than the peeling value, the intersection will be skipped over. Figure \ref{Fig_peeling} (b) gives an example of surface peeling, which contains two peeling windows.

\subsection{Voxel Selection}

\begin{figure}[!t]
\centering
    \begin{tabular}{cc}
    \includegraphics[width=1.5in]{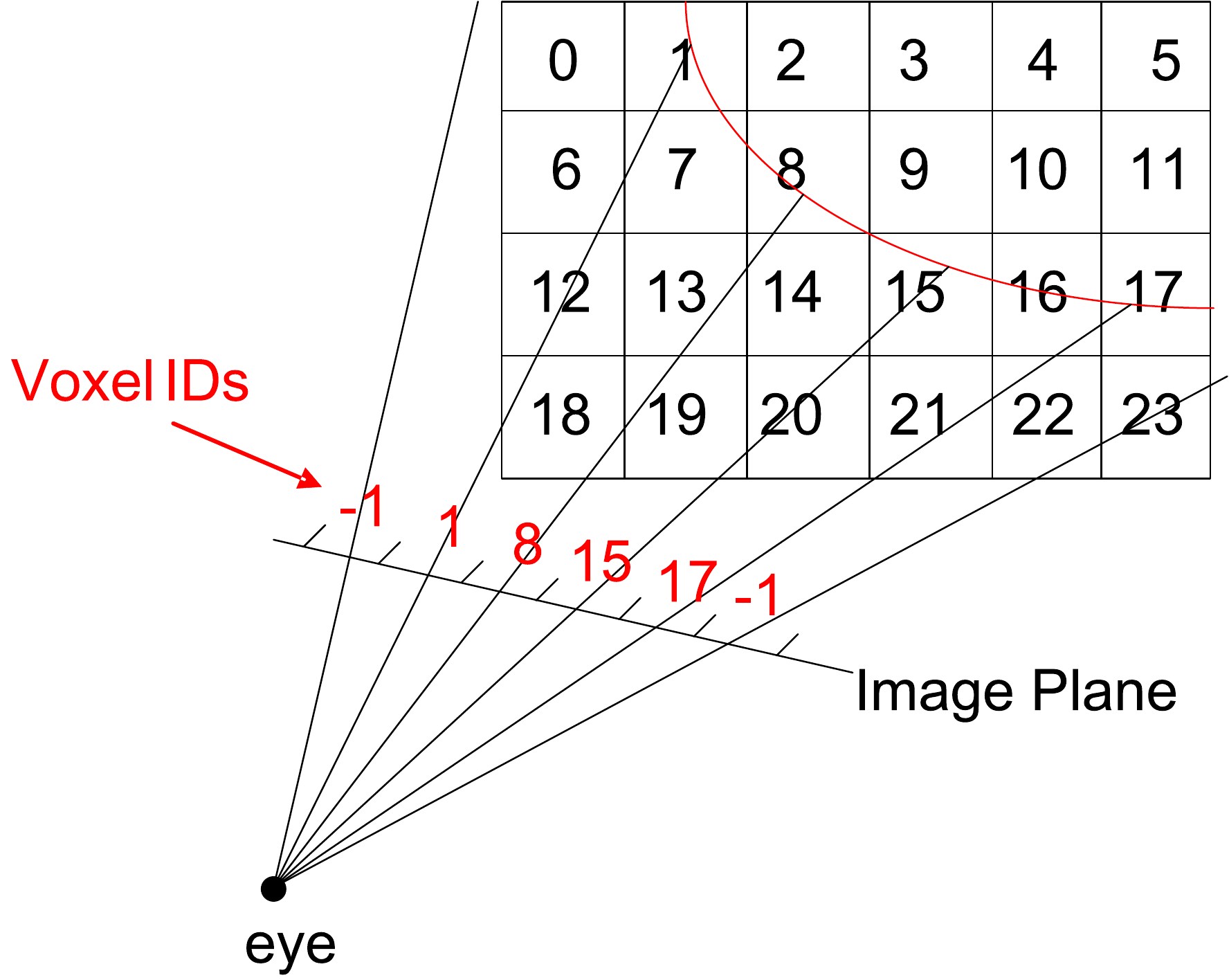}&
    \includegraphics[width=1.5in]{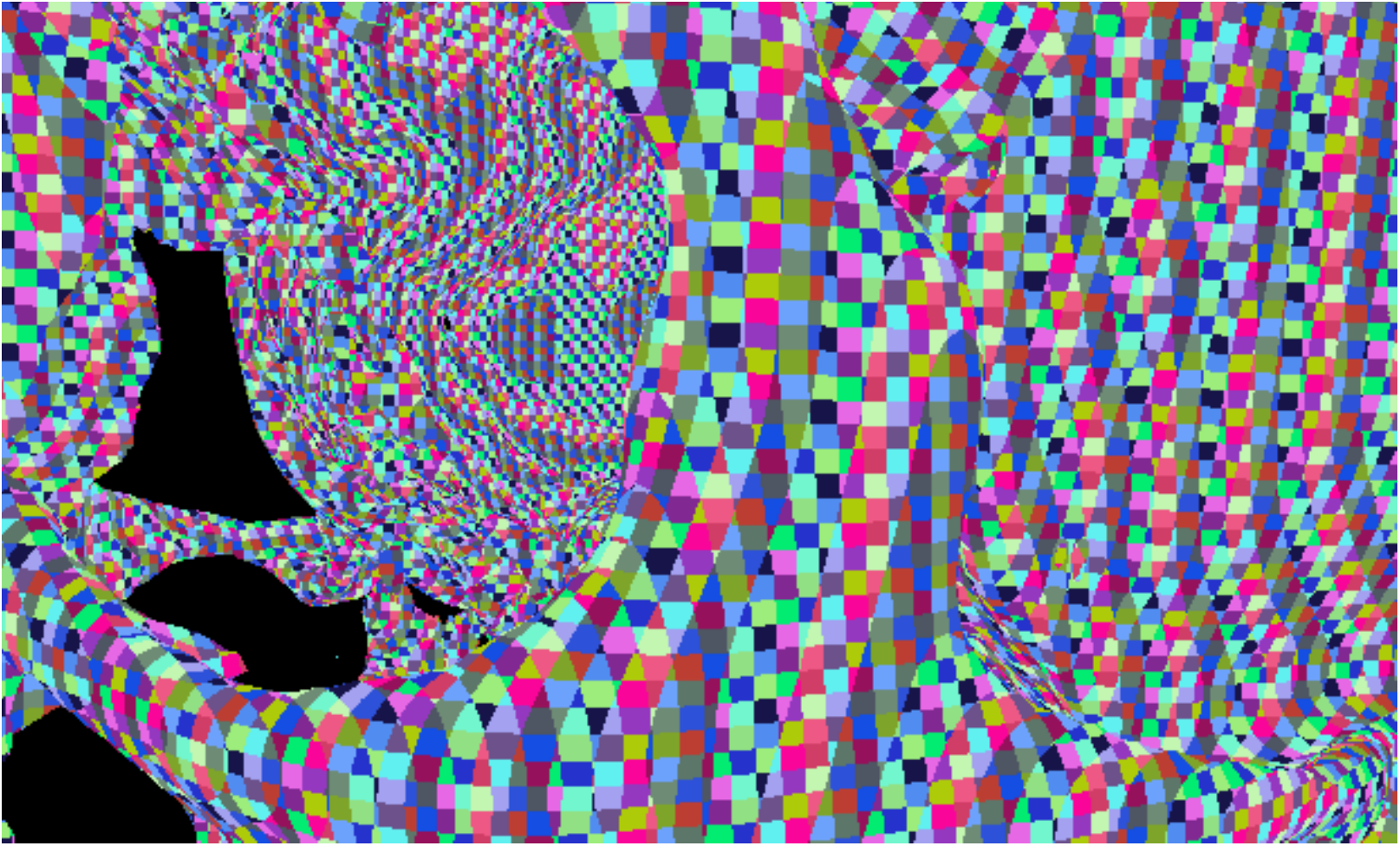}\\
    (a)&(b)\\
    \end{tabular}
\caption{Voxel Selection. (a) Illustration of the process of voxel selecting. (b) Illustration of a voxel ID buffer.} 
\label{Fig_VoxelSel}
\end{figure}

As mentioned above, when ray traversal is done in a voxel-by-voxel manner and the intersections are accurately calculated, the exact intersecting voxel can be identified for each ray intersecting the isosurface. As is shown in Figure \ref{Fig_VoxelSel} (a), a unique ID is assigned to each voxel in a sequential manner. Then, a voxel ID buffer is used to store the intersecting voxel IDs of each ray, which can be used for voxel selecting after the rendering. When user clicks or drags the mouse on the image plane, the corresponding voxel ID can be collected and organized. Figure \ref{Fig_VoxelSel} (b) illustrates a voxel ID buffer, where each small color patch is contained in a voxel. The color patches are neatly connected together due to accurate ray-isosurface intersection test.

Voxel selection in a 3D view provides a intuitive and effective interaction tool, which we used for the seed selecting in isosurface segmentation. 

\subsection{Isosurface Segmentation}

For some complicated medical datasets, segmentation is necessary for volume visualization. Traditionally, volume segmentation methods are used as the preprocessing, through which a label volume is generated. Then, rendering is done by using different transfer-functions for different volume components. These methods usually have three drawbacks. First, the segmentation is usually performed in slice views or the intensity domain, which is not intuitive. Second, the label volume, in which the segmentation result is stored has very limited precision. Stair-case artifacts can be frequently seen in rendering. Third, volume segmentation itself can be very complicated and expensive in computation. 

\begin{figure}[!t]
\centering
    \begin{tabular}{cc}
    \includegraphics[width=1.5in]{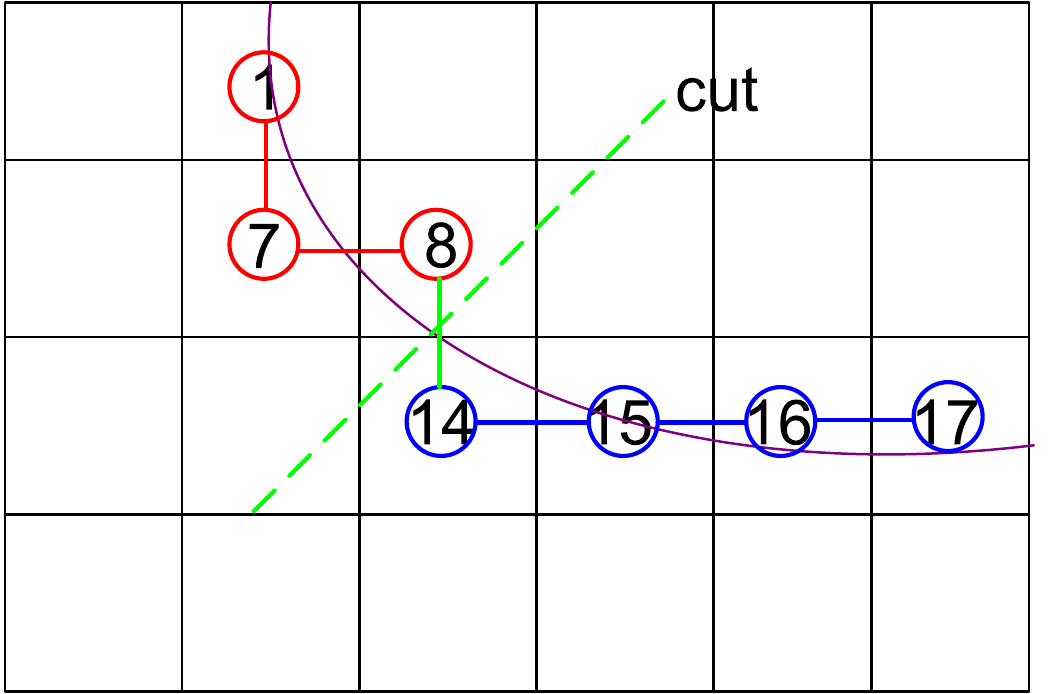}&
    \includegraphics[width=1.5in]{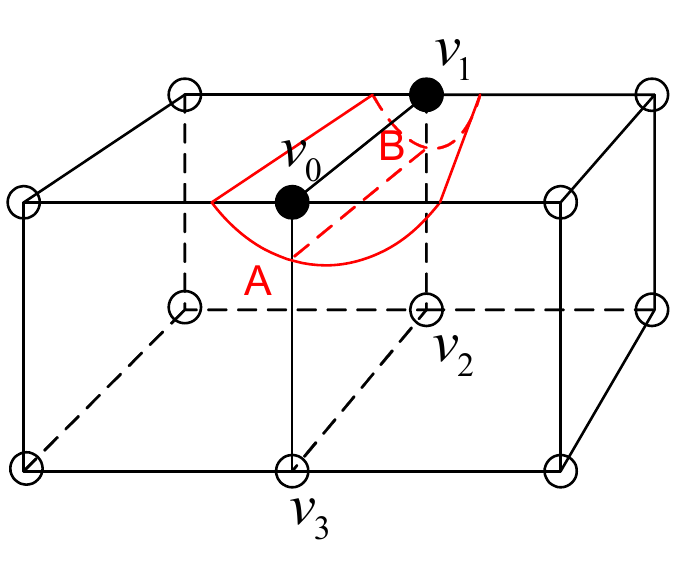}\\
    (a)&(b)\\
    \end{tabular}
\caption{Isosurface Segmentation. (a) 2D illustration of isosurface segmentation. (b) Use the length of the intersection of isosurface and voxel boundary as the weight.} 
\label{Fig_segmentation}
\end{figure}

In this part, we propose an isosurface domain segmentation method, in which we only consider the voxels containing the specific isosurface. The idea is that, these voxels can be considered as forming an inter-connected network, using 6-neighborhood, as is shown in Figure \ref{Fig_segmentation} (a), where it is 4-neighborhood in 2D. So the isosurface segmentation can be considered as a graph segmentation problem.

With the voxel selection method described above, the seed points for the foreground and the background can be directly picked from the rendering view. In our implementation, the voxel IDs of the seed points are stored as two "set" structures, which is sorted and optimized for searching. The graph is then constructed by a breadth first search from both seed sets. In the graph, the segmentation problem can be defined as to find an optimal cut that separate the two seed sets, which can be solved with the min-cut algorithm after the weight of each link is decided. For the decision of the weights, it is intuitive to consider the geometrically shortest cut as the optimal solution, since it introduces the smallest damage to the original structure. So the weight should be defined as the length of the intersection of the isosurface and the voxel boundary, as is shown in Figure \ref{Fig_segmentation} (b), the A-B segment.  

If there is only one segment of intersection of the isosurface and the voxel boundary, the length can be simply calculated by a numerical integration. However, with trilinear interpolation in volume space, there can be as many as two segments at most, although such case is rare. In that case, usually only one of the two segments is of interest during segmentation, but there's no information about which segment it is. So we treat this case by dividing the total length of the two intersection segments by 2.

During seed selection, the current selection status is previewed with a selection based coloring method. Since the seed sets are not likely to be large, they are directly transfered to the graphics memory as sequential lists. During rendering, for each intersection point, according the voxel ID information, it can be judged whether the voxel is contained in the two lists by binary searches, and the color can be decided accordingly.

With isosurface segmentation, the IDs of two sets of voxels are extracted, each containing part of the same isosurface, which we call surface structures. Such a surface structure is not guaranteed to be closed, so it cannot substitute volume segmentation. However, we will show that they can be recombined and rendered in a single scene with different coloring settings, which provides valuable visualization results. 

\subsection{Multi-surface-structure Visualization}

\begin{figure}[!t]
\centering
\includegraphics[width=2.0in] {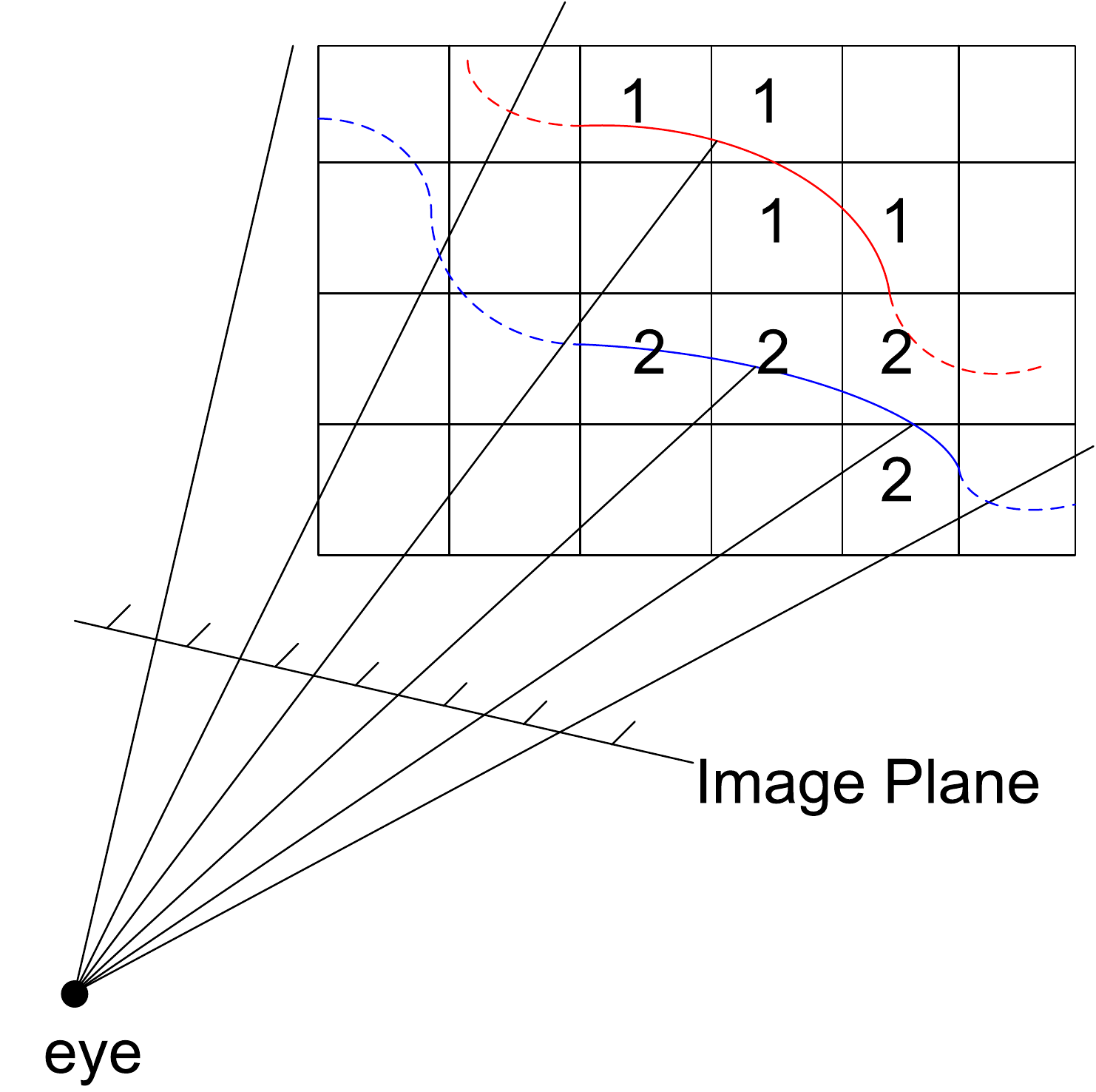}
\caption{Multi-surface-structure visualization} 
\label{Fig_MultiSurface}
\end{figure}

To combine the surface structures into a single scene, a label volume of 8-bit integer values, ranging from 0 to 255 is used. While the value 0 is used to mark the empty voxels, values from 1 to 255 are corresponding to different surface structures. So at most 255 surface structures are allowed to be defined, which is sufficient for most applications. For each surface structure, an isovalue is stored, which is the isovalue used in the isosurface segmentation stage to extract that surface structure. Also, to enable color enhancement, each surface structure has its own local transfer-function, and the speed-color maps are stored in a 2D look-up-table, where each line is corresponding to a speed-color map. 

The rendering process is just slightly different from single isosurface rendering. During ray traversal, the first step is to find a voxel with a none-zero label. Then, within the voxel, the isovalue corresponding to the label value is used for ray-isosurface intersection test, and the corresponding speed-color map is used for color enhancement. Since the ray-isosurface intersection test can still be performed at subvoxel level, the accuracy is preserved and is free of the stair case artifacts.

%% file: Experiments.tex
\section{Results}

We conducted several experiments on the proposed techniques on a PC equipped with Intel i7 950 CPU, 4GB RAM, and NVIDIA Geforce GTX 480 graphics card. The rendering computation is implemented on the GPU with CUDA. 

\subsection{Isosurface Color Enhancement}

\begin{figure*}[!t]
\centering
    \begin{tabular}{ccc}
    \includegraphics[height=2.0in]{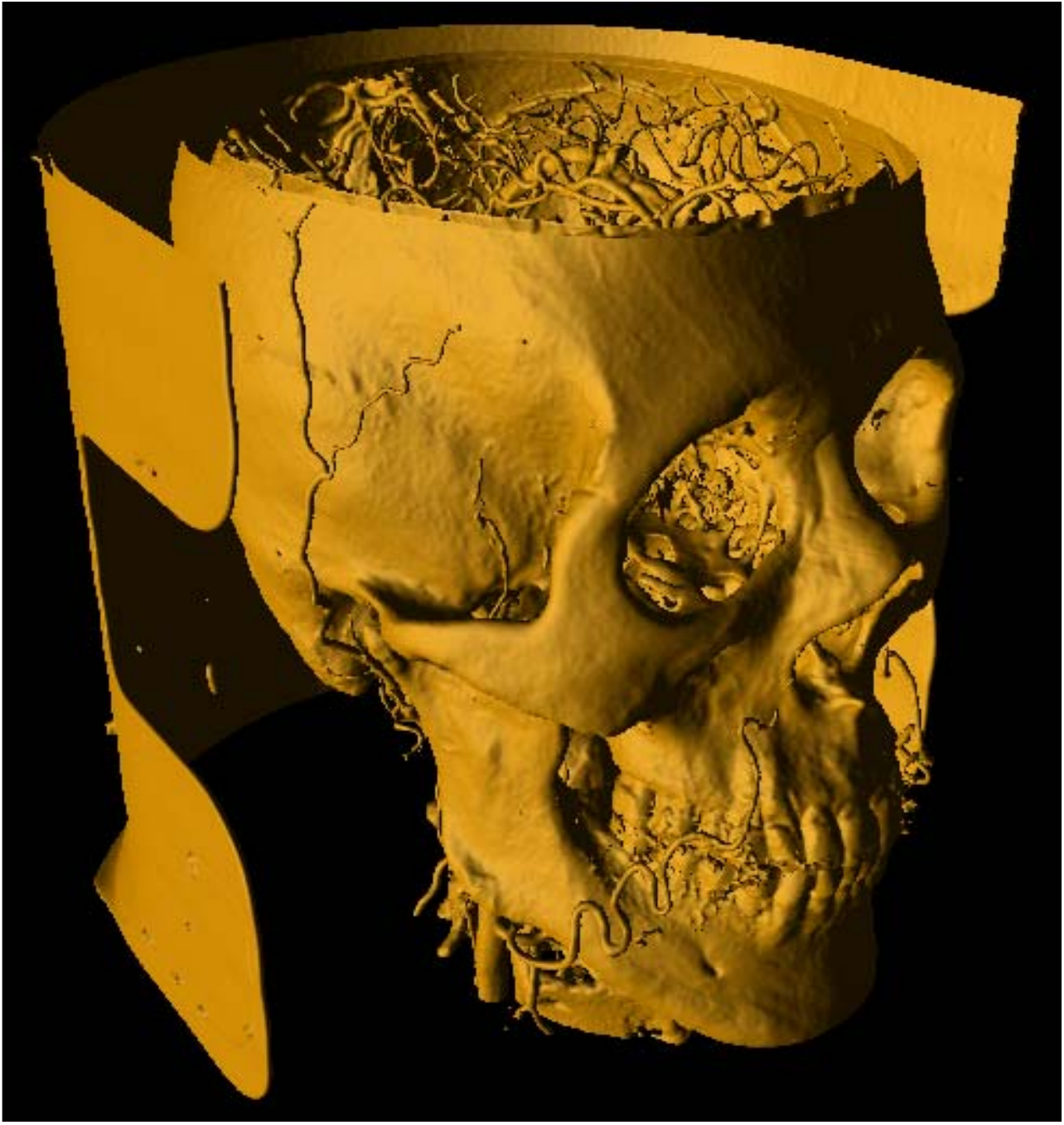}&
    \includegraphics[height=2.0in]{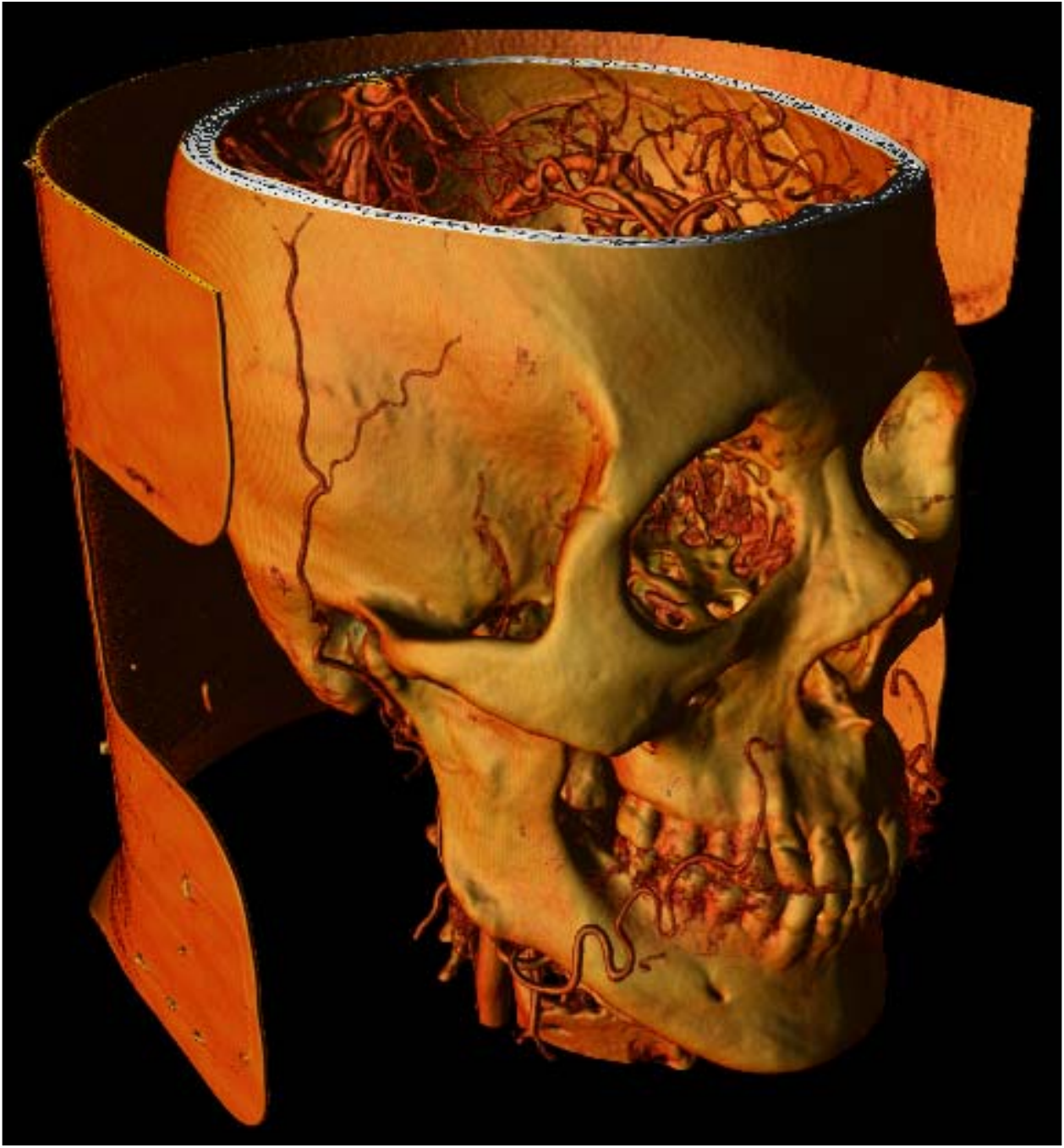}&
    \includegraphics[height=2.0in]{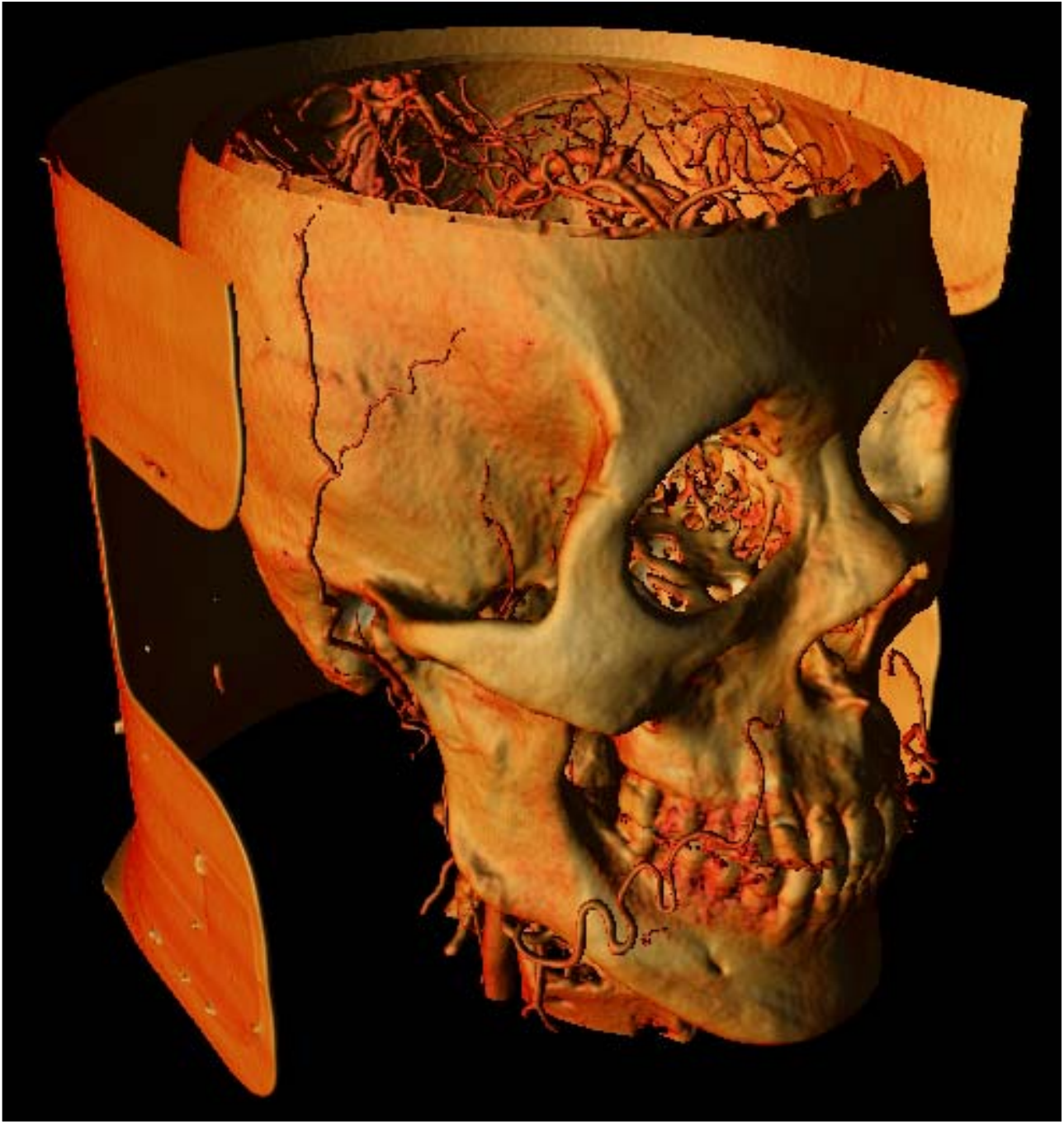}\\
    (a)&(b)&(c)\\
    \end{tabular}
\caption{Isosurface Color Enhancement Example. (a) Ordinary monotone isosurface rendering. (b) Full volume rendering. (c) Color enhanced isosurface rendering.  The size of the images is 735x556. Rendering kernel execution times per-frame: (a) 6.0834 ms (b) 21.8971 ms (c) 8.69162 ms }
\label{Fig_Iso_Color_En_Ex}
\end{figure*}

As mentioned before, isosurface color enhancement can be used in two ways due to different methods of scalar changing rate estimation. Using shallow neighborhood approximation, enhanced isosurface rendering can generate the rendering effect very similar to the full volume rendering that uses a transfer-function with a very narrow transitional section. As shown in Figure \ref{Fig_Iso_Color_En_Ex}, the visual effect of color enhanced isosurface rendering is very close to the full volume rendering, while it only takes about 2.6 more milliseconds compared with monotone isosurface rendering.

\begin{figure}[!t]
\centering
    \begin{tabular}{cc}
    \includegraphics[width=1.5in]{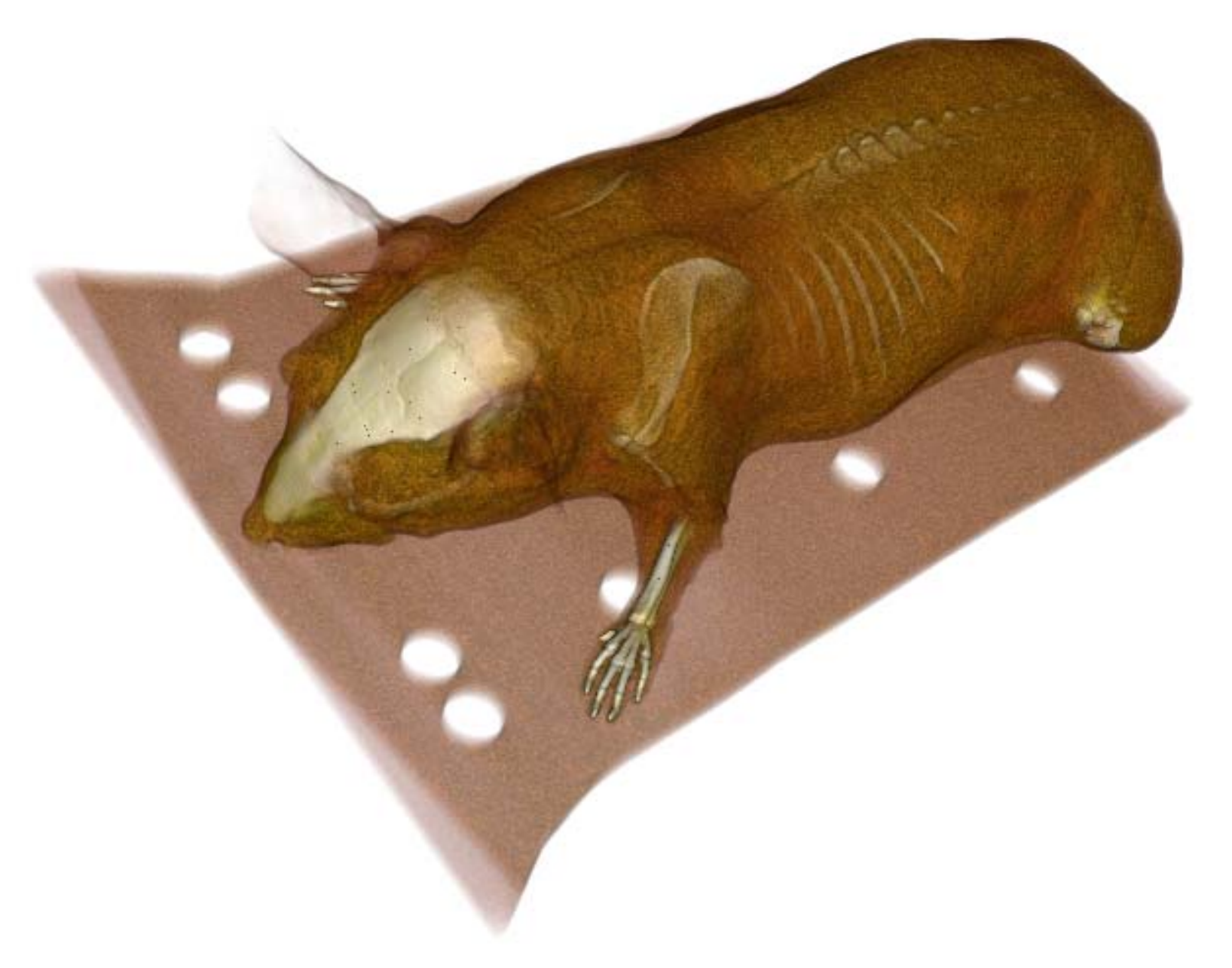}&
    \includegraphics[width=1.5in]{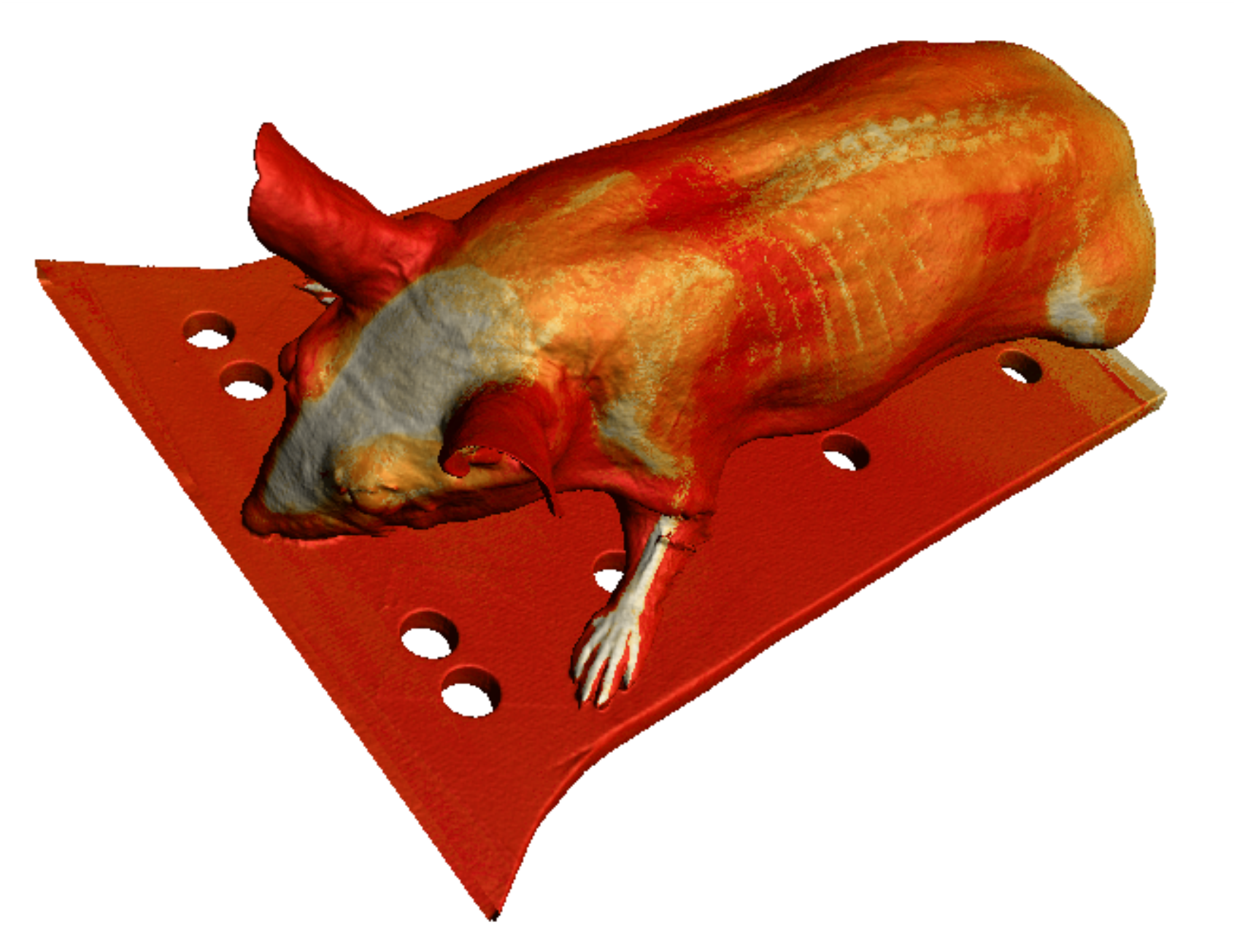}\\
    (a)&(b)\\
    \includegraphics[width=1.5in]{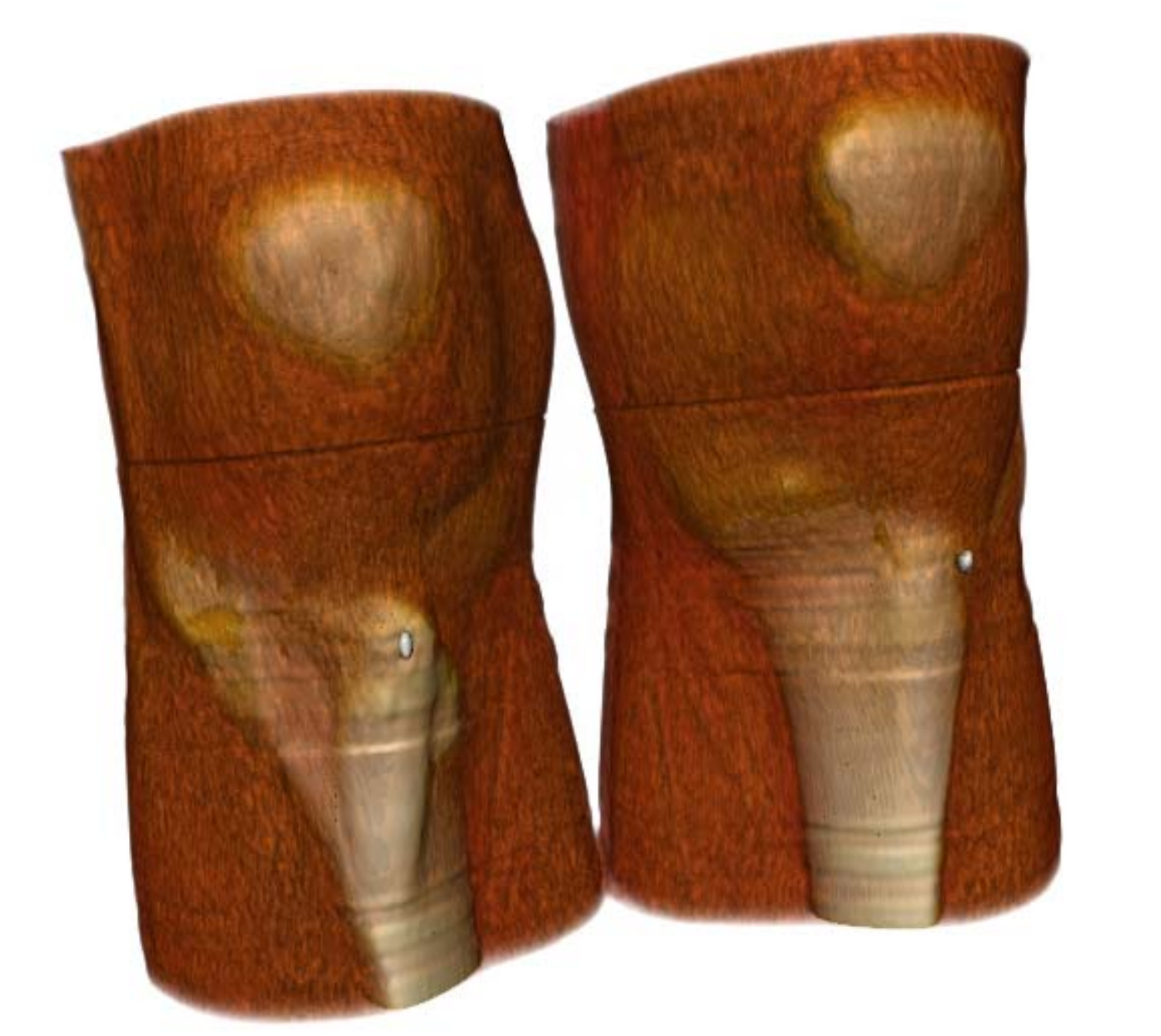}&
    \includegraphics[width=1.5in]{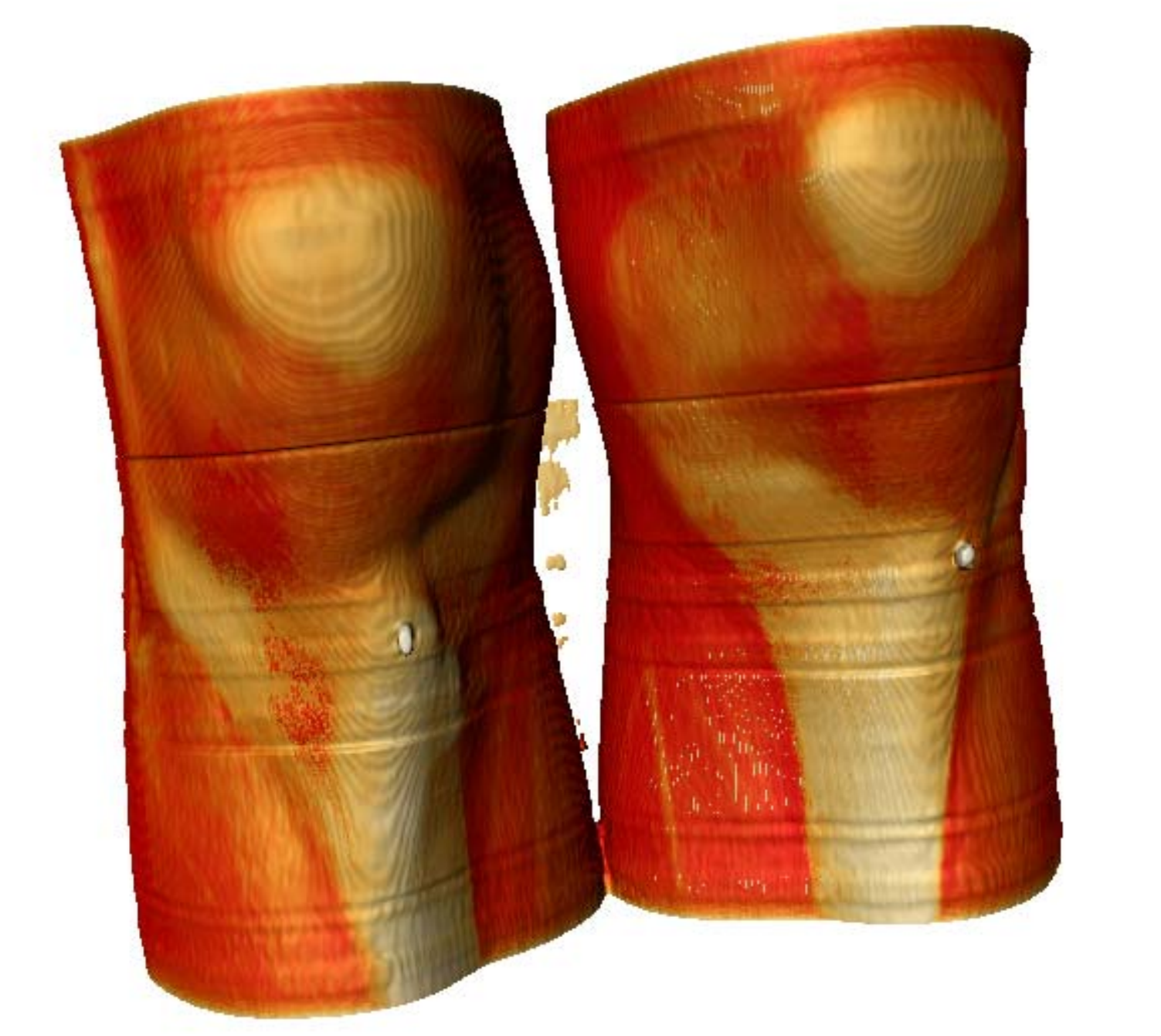}\\
    (c)&(d)\\
    \includegraphics[width=1.5in]{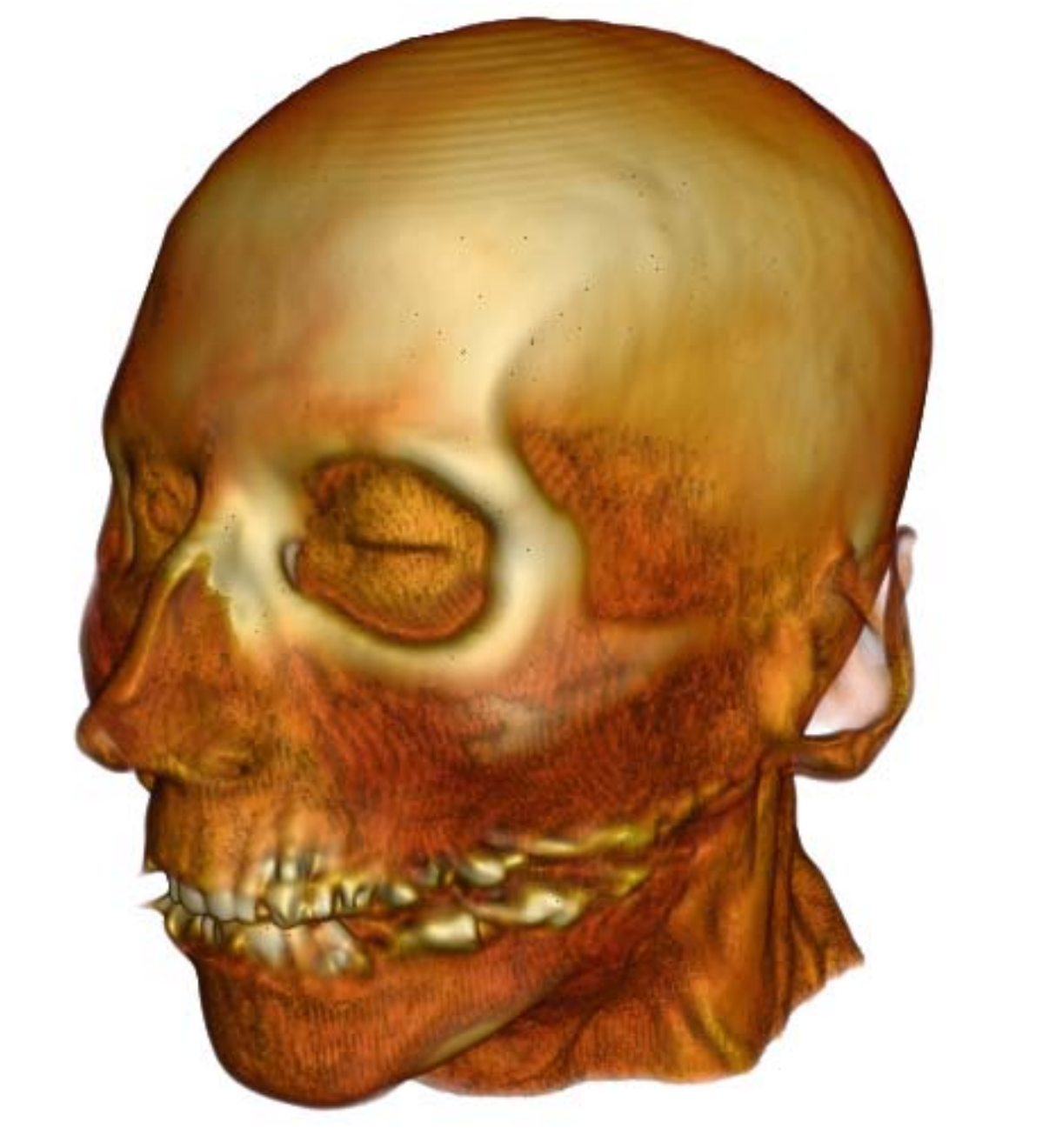}&
    \includegraphics[width=1.5in]{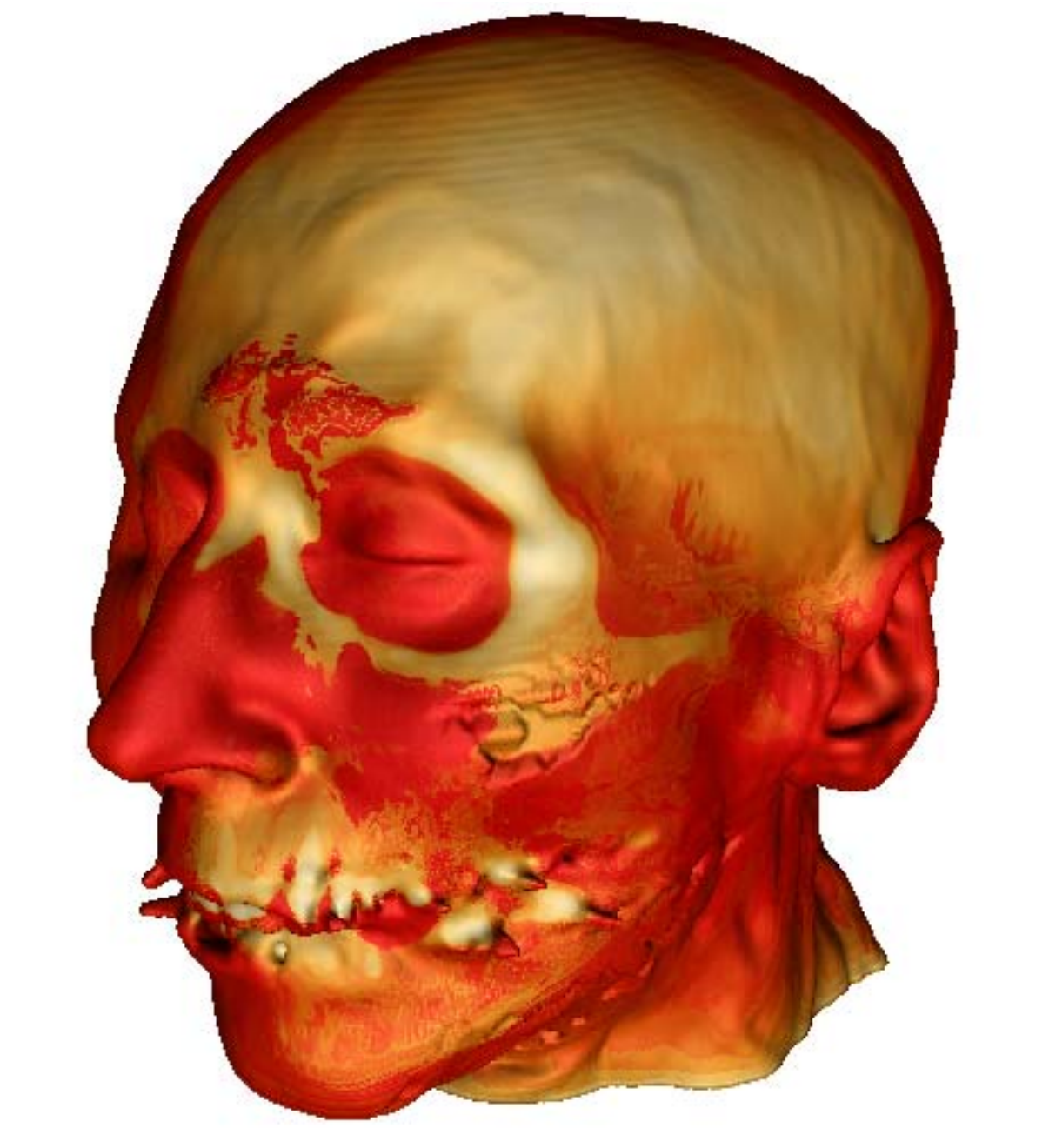}\\
    (e)&(f)\\
    \end{tabular}
\caption{Revealing internal structures. The pictures in the left column are rendered with full volume rendering(FVR) while the pictures in  the right column are rendered with color enhanced isosurface rendering (CEIR). (a)(b) The mouse data. (c)(d) The knee data. (e)(f) The head data. }
\label{Fig_Internal}
\end{figure}


\begin{table}
\centering \caption{Performance Comparison of Full Volume Rendering (FVR) and Color Enhanced Isosurface Rendering (CEIR)}
\begin{tabular}{|c|c|c|c|}
\hline Dataset & Volume Size & Image Size & \tabincell{c}{Kernel Execution \\ Time per Frame}\\
\hline Mouse & $512 \times 512 \times 512$ & $765 \times 592$ &  \tabincell{c}{FVR: 60.6232 ms \\ CEIR: 4.66121 ms } \\
\hline Knee & $379 \times 229 \times 305$ & $589 \times 488$ &  \tabincell{c}{FVR: 46.4839 ms \\ CEIR: 4.21262 ms } \\
\hline Head & $208 \times 256 \times 225$ & $713 \times 594$ &  \tabincell{c}{FVR: 34.7136 ms \\ CEIR: 4.06191 ms } \\
\hline
\end{tabular}
\label{table_performance}
\end{table}

The capability to reveal internal structures can also be compared with full volume rendering. Figure \ref{Fig_Internal} shows 3 examples of rendering the sample dataset with full volume rendering (FVR) and color enhance isosurface rendering (CEIR). Table \ref{table_performance} gives the performance comparison of FVR and CEIR. Using color enhanced isosurface rendering, while the essential structures of interest behind an isosurface can be depicted in a similar way as full volume rendering, the computational cost remains at a very low level. 

\subsection{Explicit Scene Exploration}

With the explicit scene exploration, a volume dataset can be explored in an intuitive and efficient way. It includes several techniques. Within these techniques, surface peeling enables us to explore structures deep inside the dataset. It can be applied to any isosurface rendering view. For example, it can be directly combined with isosurface color enhancement such as the example given in Figure \ref{Fig_peeling} (b). 

\begin{figure}[!t]
\centering
    \begin{tabular}{cc}
    \includegraphics[height=1.5in]{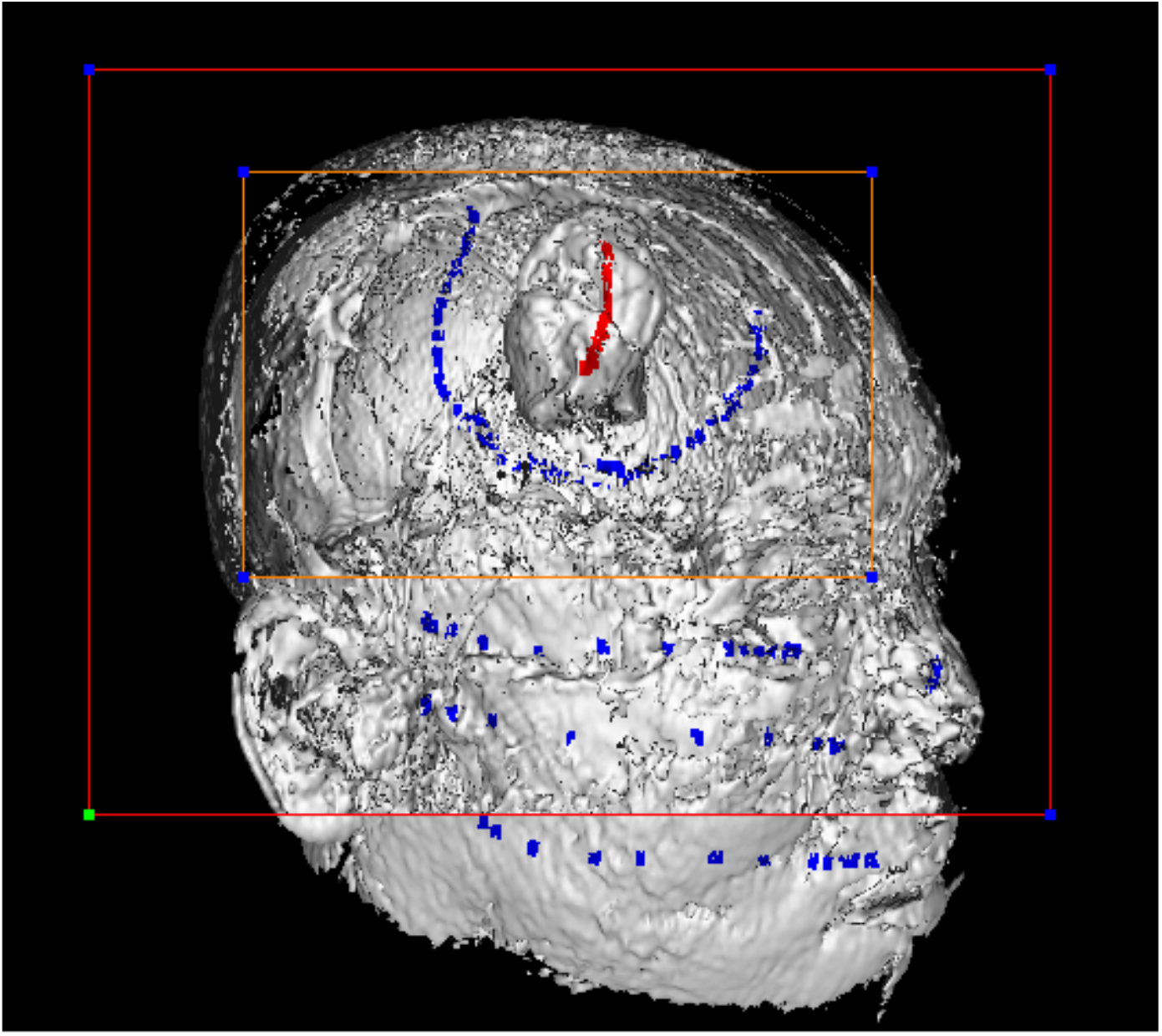}&
    \includegraphics[height=1.5in]{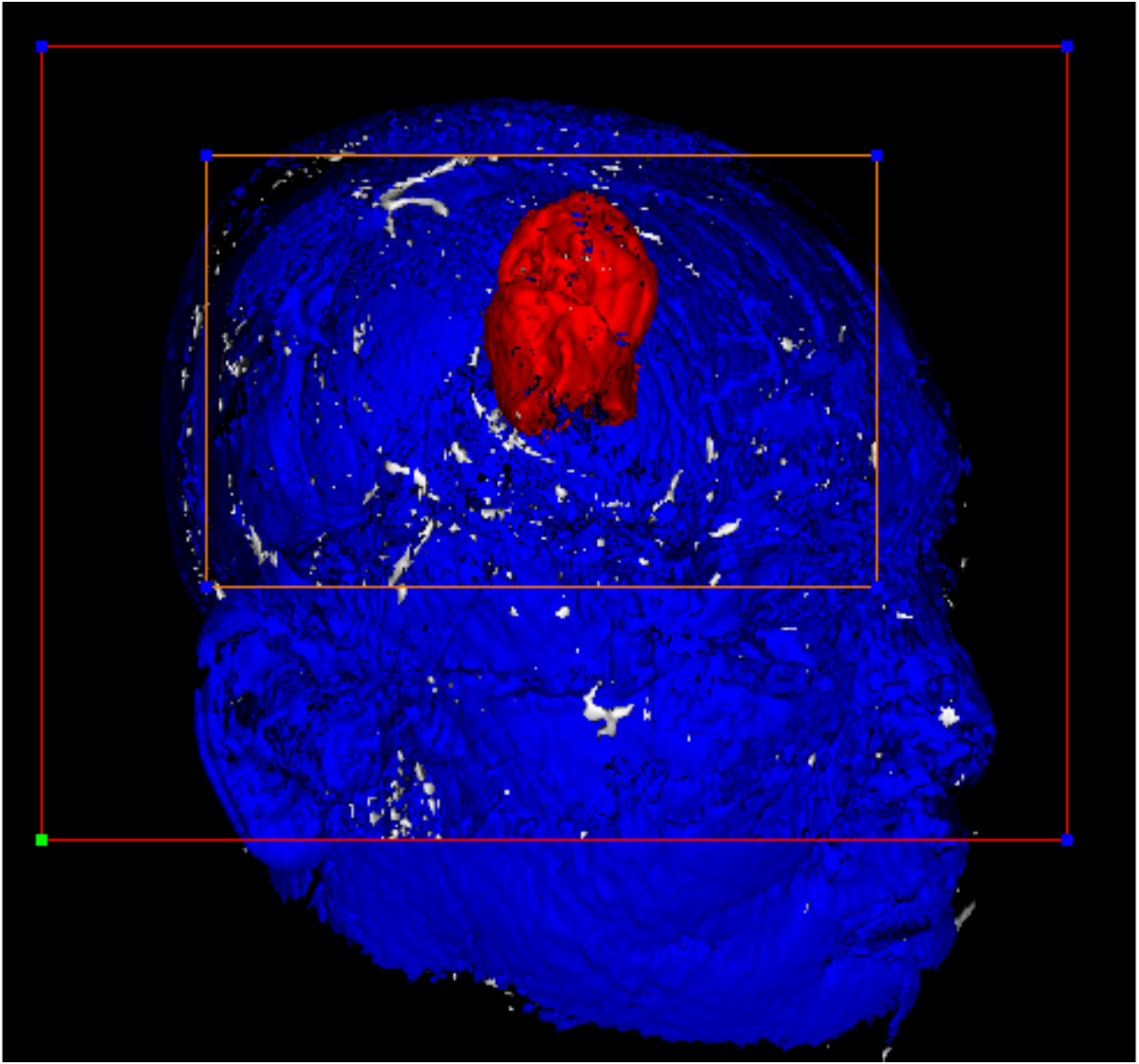}\\
    (a)&(b)\\
    \end{tabular}
\caption{Seed picking and isosurface segmentation. (a) Seed picking. (b) Isosurface segmentation. Volume size: $256 \times 256 \times 124$ Number of graph nodes: 832869 Segmentation computation time: 8.25s}
\label{Fig_Segmentation_Result}
\end{figure}

\begin{figure}[!t]
\centering
    \begin{tabular}{cc}
    \includegraphics[height=1.5in]{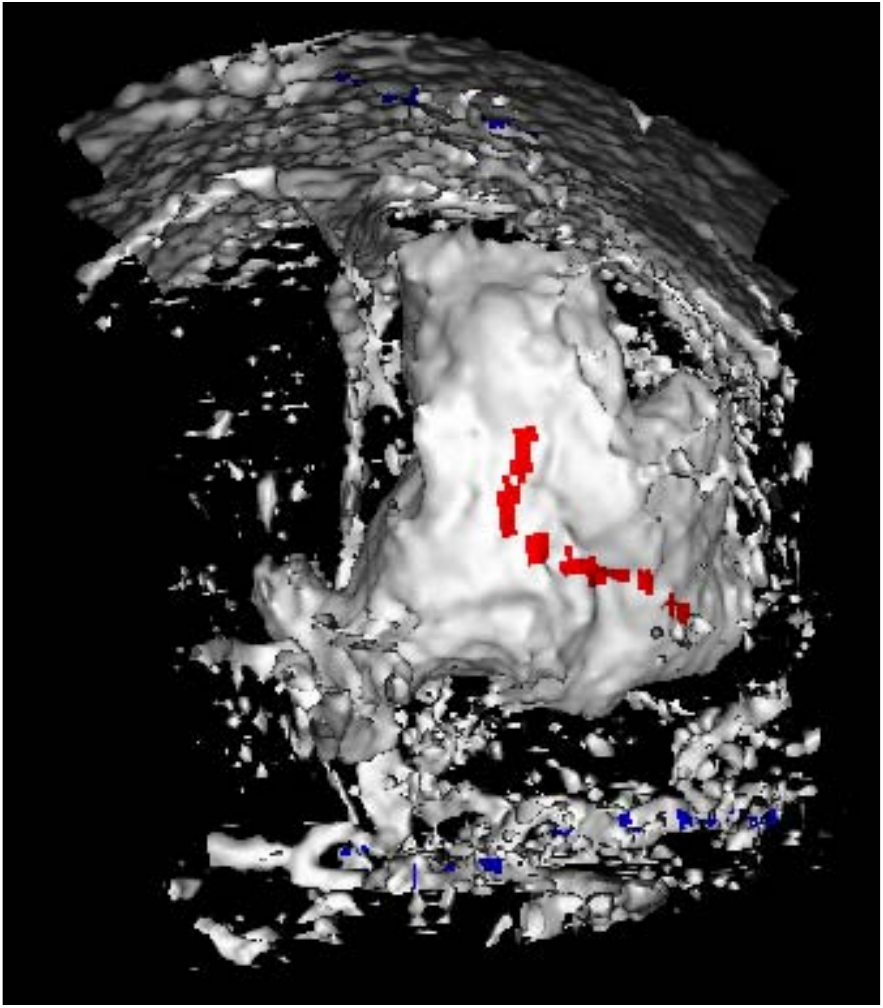}&
    \includegraphics[height=1.5in]{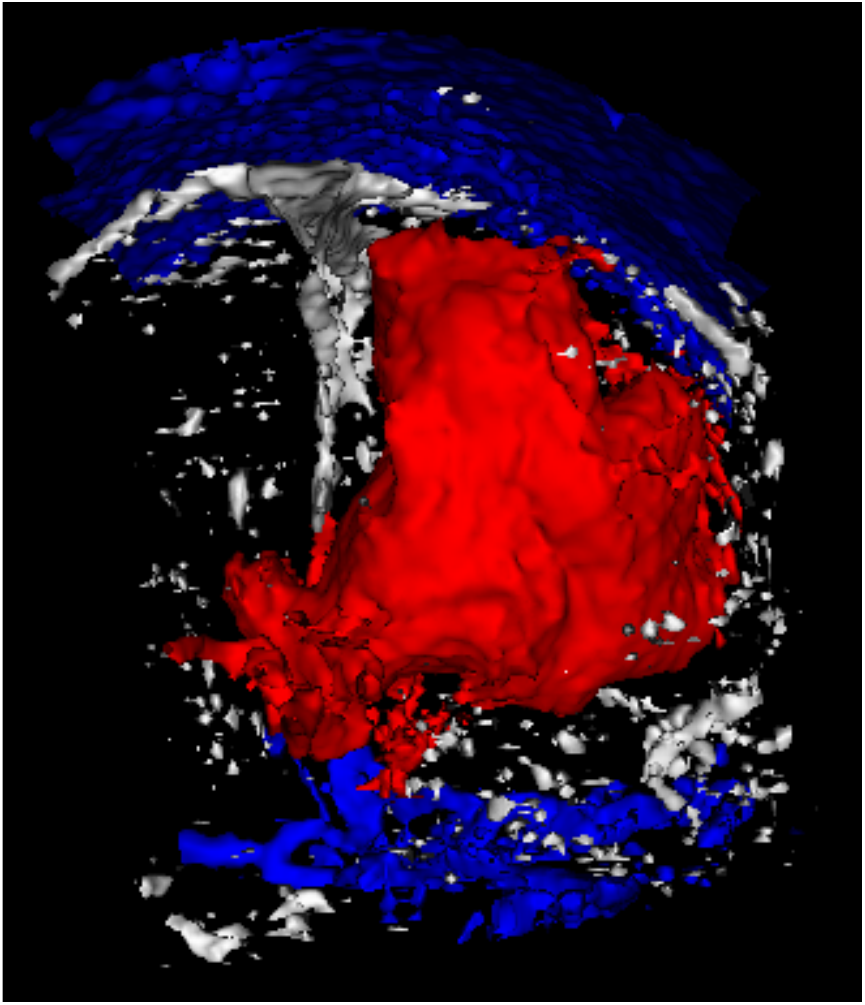}\\
    (a)&(b)\\
    \end{tabular}
\caption{Seed picking and isosurface segmentation in a cropped region. (a) Seed picking. (b) Isosurface segmentation. Number of graph nodes: 40830 Segmentation computation time: 0.36s}
\label{Fig_Crop_Seg}
\end{figure}

\begin{figure}[!t]
\centering
    \begin{tabular}{cc}
    \includegraphics[height=1.5in]{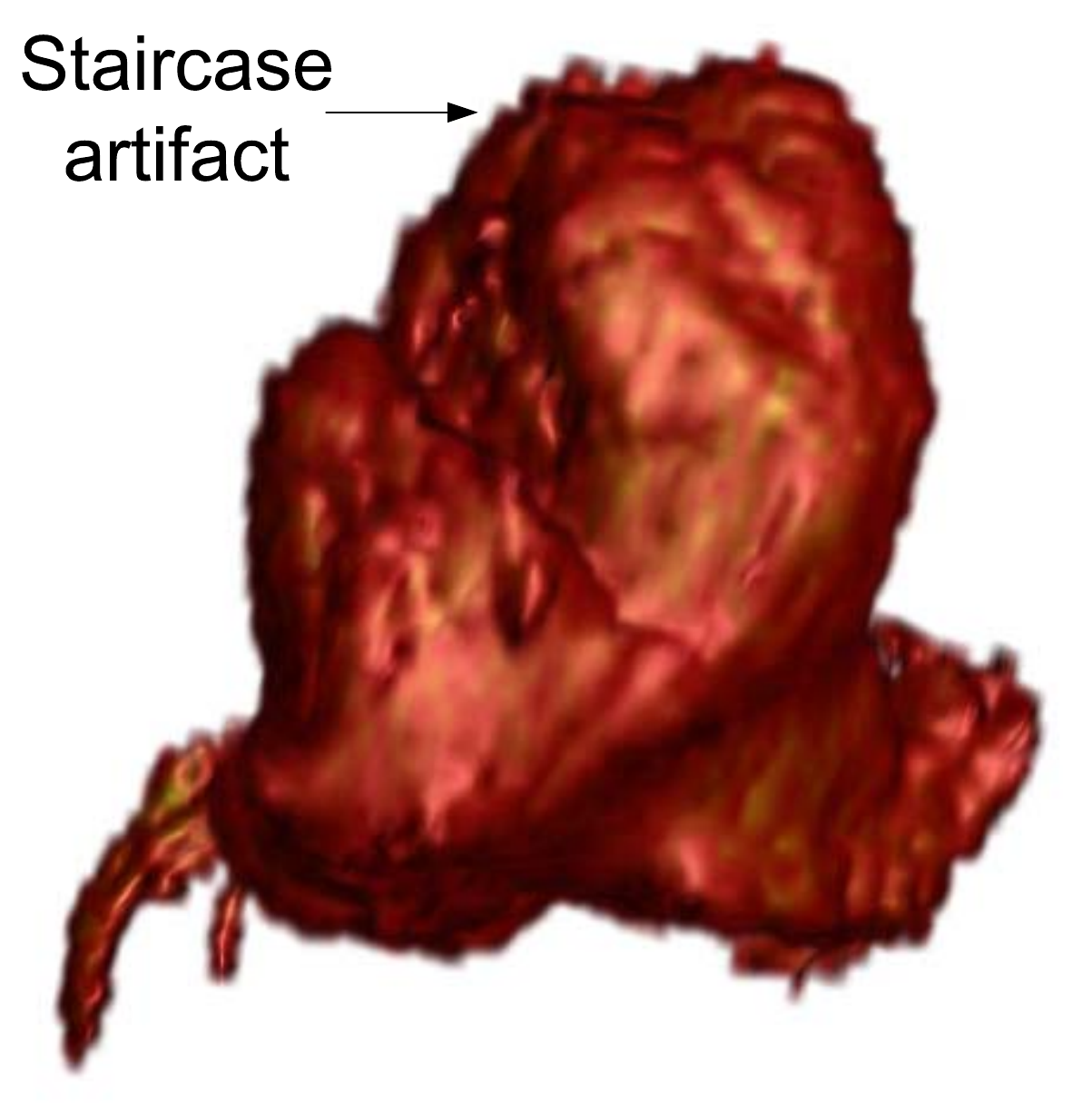}&
    \includegraphics[height=1.5in]{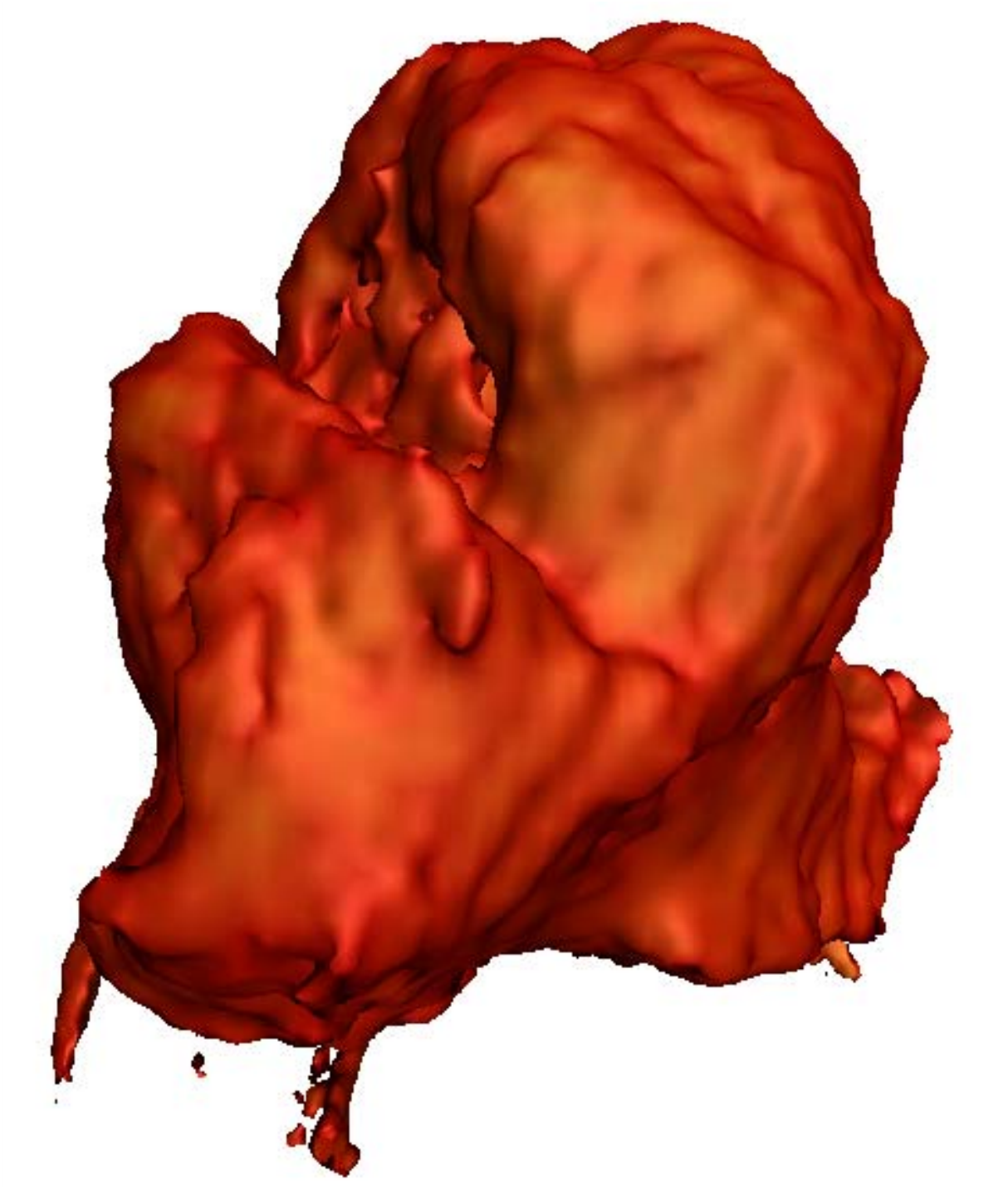}\\
    (a)&(b)\\
    \end{tabular}
\caption{Visualization accuracy comparison of volume segmentation and isosurface segmentation. (a) Rendering a segmented volume with linear boundary filtering. (b) Rendering a surface structure produces by isosurface segmentation. }
\label{Fig_Vol_vs_Sur}
\end{figure}

The other techniques mainly aim at the segmentation and recombination of the structures of interest. Figure \ref{Fig_Segmentation_Result} gives an example of seed picking and isosurface segmentation, which demonstrates how a tumor is segmented from a brain MRI image. Since the segmentation only works on the voxels containing the isosurface, the segmentation can be done very efficiently. In this example, the whole volume contains  $256 \times 256 \times 124 = 8126464$ voxels, but only $832869$ nodes are involved in the min-cut optimization, which takes 8.25s to accomplish. If cropping is applied to restrict the region of interest, the computational cost can be made even smaller. As shown in Figure \ref{Fig_Crop_Seg}, in the cropped region, the number of graph nodes is reduced to 40830, and the computing time is reduced to 0.36s. 

\begin{figure*}[!t]
\centering
\includegraphics[width=5.0in]{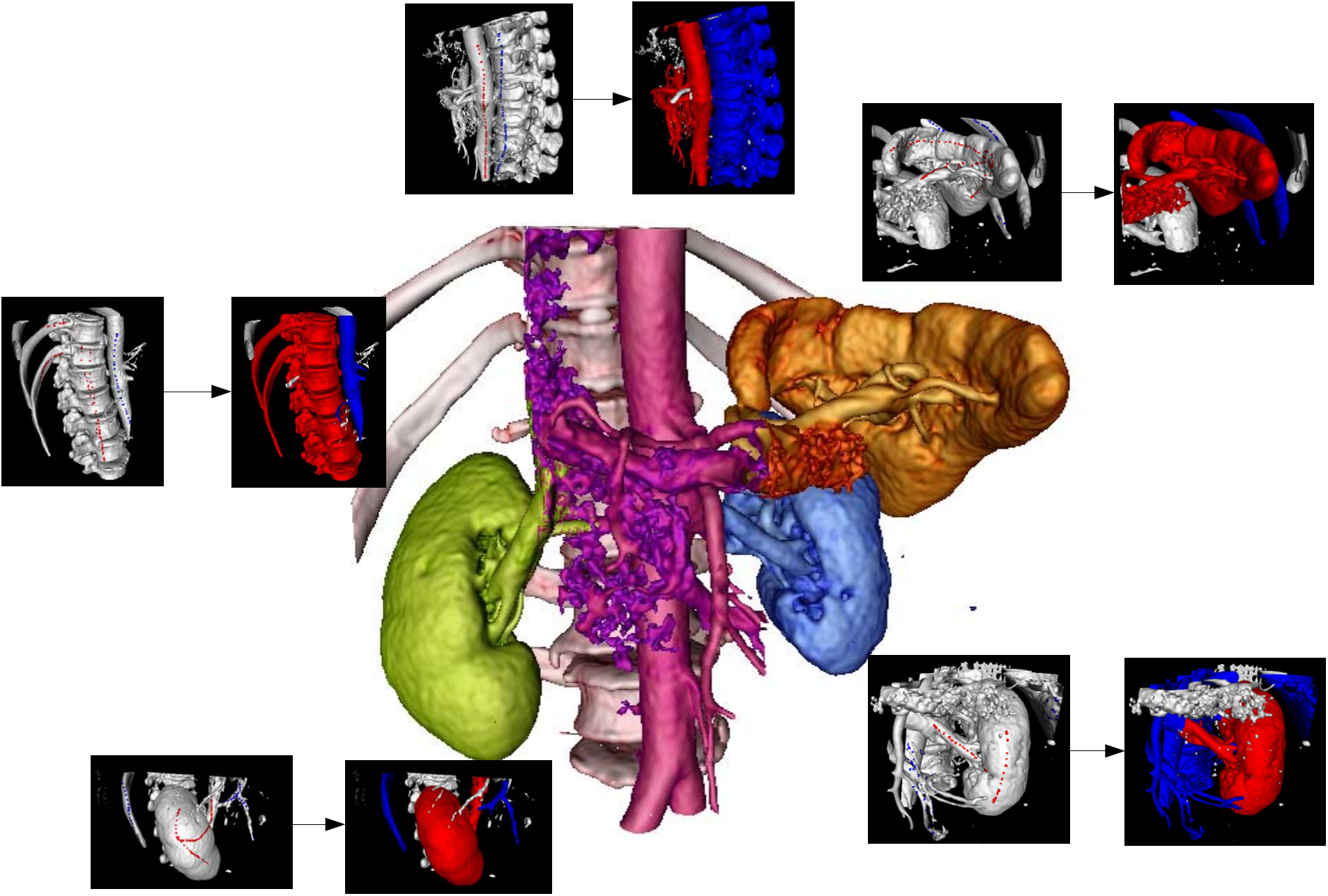}
\caption{Abdominal-structure segmentation and visualization.}
\label{Fig_Abdo_Seg_Vis}
\end{figure*}

Another benefit of isosurface segmentation is that the accurate ray-isosurface intersection can be calculated during rendering. As a result, some staircase artifacts, which are usually seen in volume segmentation based rendering, can be avoided. As shown in Figure \ref{Fig_Vol_vs_Sur} (a), if volume segmentation is used, some staircase artifacts are very likely to be introduced even if linear boundary filtering is applied. These artifacts are not seen in Figure \ref{Fig_Vol_vs_Sur} (b), where the isosurface segmentation is used.

In some applications like surgical planning, it is an essential task to extract structures of interests and to analysis the spatial relationship between them. Our explicit scene exploration provides a powerful tool to deal with these requirements. As shown in Figure \ref{Fig_Abdo_Seg_Vis}, multiple organs are segmented and represented as the surface structures, then they are recombined into a single scene for rendering.

%% file: Discussion.tex
\section{Discussion}

Among all volume visualization techniques, isosurface rendering may be the most intuitive and efficient method. It is consistent with the most widely applied optical model used in b-rep graphics, which considers objects to have a solid silhouette, and the visual features mostly lie on the surface. This model faithfully depicts the objects in reality, and the rendering result has a realistic appearance. However, for the same reason, people can argue that isosurface rendering might not have sufficient functionalities for visualizing complex volume datasets. Our work shows that such point of view is very likely to be biased. First, its intuitiveness and simplicity are the advantages for better computational performance which is critical in some interactive applications. Second, although it has some limitations, much can done to improve them and make isosurface rendering more effective. In this work, we presented a series of techniques to enhance the effectiveness of the isosurface rendering, and make it a more powerful tool suitable for more application scenarios. 

To bring more information into isosurface visualization, we propose the {\em color enhanced isosurface rendering}. Using this method, the color of isosurface is no longer monotone. It can be considered that the isosurface is painted with a texture, which is generated according the neighborhood density, in order to reveal the internal structures of the volume data. As such, the representative power of isosurface rendering is very much extended. At the same time, its theoretical groundings can be traced to the full volume rendering. 

The explicit scene exploration further extends isosurface rendering by allowing physically meaningful structures to be extracted from different isosurfaces. In contrast to direct volume segmentation, the isosurface segmentation method can be more seamlessly integrated to an isosurface rendering system. However, since the extracted surface is not guaranteed to be closed, it can't be used to replace volume segmentation. Despite of this, we hope to extend the segmentation method to volume segmentation by closing the cuts generated in the segmentation. In this way, the applications of the technique may be extended to requirements other than visualization, such as automated structure analysis.

%% file: paper.bbl
\begin{thebibliography}{\protect\citename{Stegmaier et~al\mbox{.} }2005}

\bibitem[\protect\citename{Amanatides and Woo }1987]{Amanatides:1987:AFV}
{\sc Amanatides, J., and Woo, A.}
\newblock 1987.
\newblock A fast voxel traversal algorithm for ray tracing.
\newblock In {\em In Eurographics ¡¯87}, 3--10.

\bibitem[\protect\citename{Ament et~al\mbox{.} }2010]{Ament:2010:DIVV}
{\sc Ament, M., Weiskopf, D., and Carr, H.}
\newblock 2010.
\newblock Direct interval volume visualization.
\newblock {\em Visualization and Computer Graphics, IEEE Transactions on 16},
  6, 1505 --1514.

\bibitem[\protect\citename{Boykov and Kolmogorov }2004]{boykov:2004:exp}
{\sc Boykov, Y., and Kolmogorov, V.}
\newblock 2004.
\newblock An experimental comparison of min-cut/max-flow algorithms for energy
  minimization in vision.
\newblock {\em Pattern Analysis and Machine Intelligence, IEEE Transactions on
  26}, 9, 1124--1137.

\bibitem[\protect\citename{Caban and Rheingans }2008]{caban:2008:texture}
{\sc Caban, J., and Rheingans, P.}
\newblock 2008.
\newblock Texture-based transfer functions for direct volume rendering.
\newblock {\em Visualization and Computer Graphics, IEEE Transactions on 14},
  6, 1364--1371.

\bibitem[\protect\citename{Correa and Ma }2008]{correa:2008:size}
{\sc Correa, C., and Ma, K.}
\newblock 2008.
\newblock Size-based transfer functions: A new volume exploration technique.
\newblock {\em Visualization and Computer Graphics, IEEE Transactions on 14},
  6, 1380--1387.

\bibitem[\protect\citename{Engel et~al\mbox{.} }2001]{engel:2001:HPV}
{\sc Engel, K., Kraus, M., and Ertl, T.}
\newblock 2001.
\newblock {High-quality pre-integrated volume rendering using
  hardware-accelerated pixel shading}.
\newblock In {\em Proceedings of the ACM SIGGRAPH/EUROGRAPHICS workshop on
  Graphics hardware}, ACM New York, NY, USA, 9--16.

\bibitem[\protect\citename{Hadwiger et~al\mbox{.} }2003]{hadwiger:2003:HTV}
{\sc Hadwiger, M., Berger, C., and Hauser, H.}
\newblock 2003.
\newblock High-quality two-level volume rendering of segmented data sets on
  consumer graphics hardware.
\newblock In {\em Visualization, 2003. VIS 2003. IEEE}, IEEE, 301--308.

\bibitem[\protect\citename{Kniss et~al\mbox{.} }2002]{kniss:2002:MTF}
{\sc Kniss, J., Kindlmann, G., and Hansen, C.}
\newblock 2002.
\newblock Multidimensional transfer functions for interactive volume rendering.
\newblock {\em Visualization and Computer Graphics, IEEE Transactions on 8}, 3,
  270--285.

\bibitem[\protect\citename{Lum et~al\mbox{.} }2004]{lum:2004:HLA}
{\sc Lum, E., Wilson, B., and Ma, K.}
\newblock 2004.
\newblock {High-quality lighting and efficient pre-integration for volume
  rendering}.
\newblock In {\em Proceedings Joint Eurographics-IEEE TVCG Symposium on
  Visualization 2004 (VisSym¡¯04)}, 25--34.

\bibitem[\protect\citename{Marmitt et~al\mbox{.} }2004]{Marmitt:2004:FAA}
{\sc Marmitt, G., Kleer, A., Wald, I., and Friedrich, H.}
\newblock 2004.
\newblock Fast and accurate ray-voxel intersection techniques for iso-surface
  ray tracing.
\newblock In {\em in Proceedings of Vision, Modeling, and Visualization (VMV)},
  429--435.

\bibitem[\protect\citename{Max }1995]{Max:1995:OMF}
{\sc Max, N.}
\newblock 1995.
\newblock Optical models for direct volume rendering.
\newblock {\em IEEE Transactions on Visualization and Computer Graphics 1}, 2
  (June), 99--108.

\bibitem[\protect\citename{Neubauer et~al\mbox{.} }2002]{Neubauer:2002:CFR}
{\sc Neubauer, A., Mroz, L., Hauser, H., and Wegenkittl, R.}
\newblock 2002.
\newblock Cell-based first-hit ray casting.
\newblock In {\em Proceedings of the symposium on Data Visualisation 2002},
  Eurographics Association, Aire-la-Ville, Switzerland, Switzerland, VISSYM
  '02, 77--ff.

\bibitem[\protect\citename{Parker et~al\mbox{.} }1998]{Parker:1998:IRT}
{\sc Parker, S., Shirley, P., Livnat, Y., Hansen, C., and Sloan, P.-P.}
\newblock 1998.
\newblock Interactive ray tracing for isosurface rendering.
\newblock {\em Visualization Conference, IEEE 0\/}, 233.

\bibitem[\protect\citename{Pickhardt }2004]{Pickhardt:2004:translucency}
{\sc Pickhardt, P.}
\newblock 2004.
\newblock Translucency rendering in 3d endoluminal ct colonography: a useful
  tool for increasing polyp specificity and decreasing interpretation time.
\newblock {\em American Journal of Roentgenology 183}, 2, 429--436.

\bibitem[\protect\citename{Scharsach }2005]{Scharsach:2005:AGR}
{\sc Scharsach, H.}
\newblock 2005.
\newblock Advanced gpu raycasting.
\newblock In {\em the 9th Central European Seminar on Computer Graphics}.

\bibitem[\protect\citename{Stegmaier et~al\mbox{.} }2005]{Stemaier:2005:SFVR}
{\sc Stegmaier, S., Strengert, M., Klein, T., and Ertl, T.}
\newblock 2005.
\newblock A simple and flexible volume rendering framework for
  graphics-hardware-based raycasting.
\newblock {\em International Workshop on Volume Graphics 0\/}, 187--241.

\bibitem[\protect\citename{Xiang et~al\mbox{.} }2010]{xiang:2010:SCE}
{\sc Xiang, D., Tian, J., Yang, F., Yang, Q., Zhang, X., Li, Q., and Liu, X.}
\newblock 2010.
\newblock Skeleton cuts-an efficient segmentation method for volume rendering.
\newblock {\em Visualization and Computer Graphics, IEEE Transactions on}, 99,
  1--1.

\end{thebibliography}
